\documentclass[Journal]{IEEEtran}

\usepackage{amsmath,amssymb,float,arydshln,color}
\usepackage{amsthm}
\usepackage{cite}
\usepackage{psfrag,setspace,wrapfig,subfigure}
\usepackage[latin1]{inputenc}
\usepackage{dsfont}
\usepackage{epsfig}
\usepackage{epstopdf}
\usepackage{graphicx}
\usepackage{hyperref}
\usepackage{amsfonts}
\usepackage{bbm}
\usepackage{tikz}
\usetikzlibrary{fit}
\usepackage{empheq}
\newcommand*\widefbox[1]{\fbox{\hspace{2em}#1\hspace{2em}}}
\allowdisplaybreaks

\tikzset{%
	highlight/.style={circle,draw,minimum size=0.1cm,inner sep=0pt}
}

\newtheorem{lemma}{{Lemma}}
\newtheorem*{assumption*}{{ Assumption}}
\newtheorem{theorem}{{Theorem}}

\def\tran{^{\mathsf{T}}}
\def\one{\mathds{1}}

\newcommand{\bp}{\small \begin{proof}}
	\newcommand{\ep}{\end{proof} \normalsize}

\newcommand{\bm}[1]{\mbox{\boldmath $#1$}}

\newcommand{\be}{\begin{equation}}
	\newcommand{\ee}{\end{equation}}
\newcommand{\bal}{\begin{align}}
	\newcommand{\eal}{\end{align}}
\newcommand{\bq}{\begin{eqnarray}}
	\newcommand{\eq}{\end{eqnarray}}
\newcommand{\bqn}{\begin{eqnarray*}}
	\newcommand{\eqn}{\end{eqnarray*}}
\newcommand{\nn}{\nonumber}
\newcommand{\ba}{\left[ \begin{array}}
	\newcommand{\ea}{\\ \end{array} \right]}
\newcommand{\qd}{\hfill{$\blacksquare$}}
\newcommand{\define}{\;\stackrel{\Delta}{=}\;}
\newcommand{\aseq}{\;\stackrel{a.s.}{=}\;}

\def\bmu  	{{\boldsymbol \mu}}

\def\t{{\boldsymbol{t}}}

\def\Zint{{\mathchoice{\setbox1=\hbox{\sf Z}\copy1\kern-.75\wd1\box1}
		{\setbox1=\hbox{\sf Z}\copy1\kern-.75\wd1\box1}
		{\setbox1=\hbox{\scriptsize\sf Z}\copy1\kern-.75\wd1\box1}
		{\setbox1=\hbox{\scriptsize\sf Z}\copy1\kern-.75\wd1\box1}}}

\makeatletter
\def\hlinewd#1{%
	\noalign{\ifnum0=`}\fi\hrule \@height #1 \futurelet
	\reserved@a\@xhline}
\makeatother

\begin{document}
\def\helvetica{phvr7t.tfm}
\def\helveticaoblique{phvro7t.tfm}
\def\helveticabold{phvb7t.tfm}
\def\helveticaboldoblique{phvbo7t.tfm}

\font\sfb=\helveticabold
=\helveticaboldoblique
\title{Belief Control Strategies for Interactions \\over Weakly-Connected Graphs}
	\author{Hawraa~Salami,~\IEEEmembership{Student Member,~IEEE,}
		Bicheng~Ying,~\IEEEmembership{Student Member,~IEEE,}\\
		and~Ali~H.~Sayed,~\IEEEmembership{Fellow,~IEEE}
		
		\thanks{A short version of this work appears in the conference publication \cite{icassp2HS}.}
		\thanks{This work was supported in part by NSF grants CCF-1524250 and ECCS-
			1407712. A. H. Sayed is with the \'{E}cole Polytechnique F\'{e}d\'{e}rale de Lausanne, EPFL,   School of Engineering, CH-1015 Lausanne, Switzerland. H. Salami and B. Ying are with the Department of Electrical Engineering, University of California, Los Angeles, CA 90025. Emails: \{hsalami, ybc\}@ucla.edu and ali.sayed@epfl.ch}
	}
\maketitle
\small
\begin{abstract}
	In diffusion social learning over weakly-connected graphs, it has been shown recently that influential agents shape the beliefs of non-influential agents. This paper analyzes this mechanism more closely and addresses two main questions. First, the article examines how much freedom influential agents have in controlling the beliefs of the receiving agents, namely, whether receiving agents can be driven
	to arbitrary beliefs and whether the network structure limits the scope of
	control by the influential agents. Second, even if there is a limit to
	what influential agents can accomplish, this article develops mechanisms by which they can lead receiving agents to adopt certain beliefs. These questions raise interesting possibilities about belief control over networked agents. Once addressed, one ends up with design procedures that
	allow influential agents to drive other agents to endorse particular
	beliefs regardless of their local observations or convictions. The theoretical findings are illustrated by means of examples.
	
\end{abstract}
\section{Introduction and Motivation}
Several studies have examined the propagation of information over social networks and the influence of the graph topology on this dynamics \cite{zhao2012learning,jadbabaie2012non,Acemoglu,Yildiz,OpinionDynamic,misinformation,jackson,molavi2013reaching,sca,sca2,krishna1,Vika1,Vika2,Vika3,rib,kar2,latiha,epstein2010non,Liu2014,Su2016,8015179,7891016,80,6315271,6930814,ying2014information,icassp2}. In recent works \cite{ying2014information,icassp2,salamijournal}, an intriguing phenomenon was revealed whereby it was shown that weakly-connected graphs enable certain agents to control the opinion of other agents to great degree, irrespective of the observations sensed by these latter agents. For example, agents can be made to believe that it is ``raining'' while they happen to be observing ``sunny conditions''. Weak graphs arise in many contexts, including in popular social platforms like Twitter and similar online tools. In these graphs, the topology consists of multiple sub-networks where at least one sub-network (called a sending sub-network) feeds information in one direction to other network components without receiving back (or being interested in) any information from them. For example, a celebrity user in Twitter may have a large number of followers (running into the millions), while the user himself may not be tracking or following any (or only a small fraction) of these users. For such networks with weak graphs, it was shown in \cite{icassp2,salamijournal} that, irrespective of the local observations sensed by the receiving agents, a sending sub-network plays a domineering role and influences the beliefs of the other groups in a significant manner. In particular, receiving agents can be made to arrive at incorrect inference decisions; they can also be made to disagree on their inferences among themselves.

The purpose of this article is to examine this dynamics more closely and to reveal new critical properties, including the development of control mechanisms. We have three main contributions. First, we show that the internal graph structure connecting the receiving agents imposes a form of resistance to manipulation, but only to a certain degree. Second, we characterize the set of states that can be imposed on receiving networks; while this set is large, it turns out that it is not unlimited. And, third, for any attainable state, we develop a control mechanism that allows sending agents to force the receiving agents to reach that state and behave in that manner. 

\subsection{Weakly-Connected Graphs}
\begin{figure}[h!]
    \centering
	\includegraphics[scale=0.35]{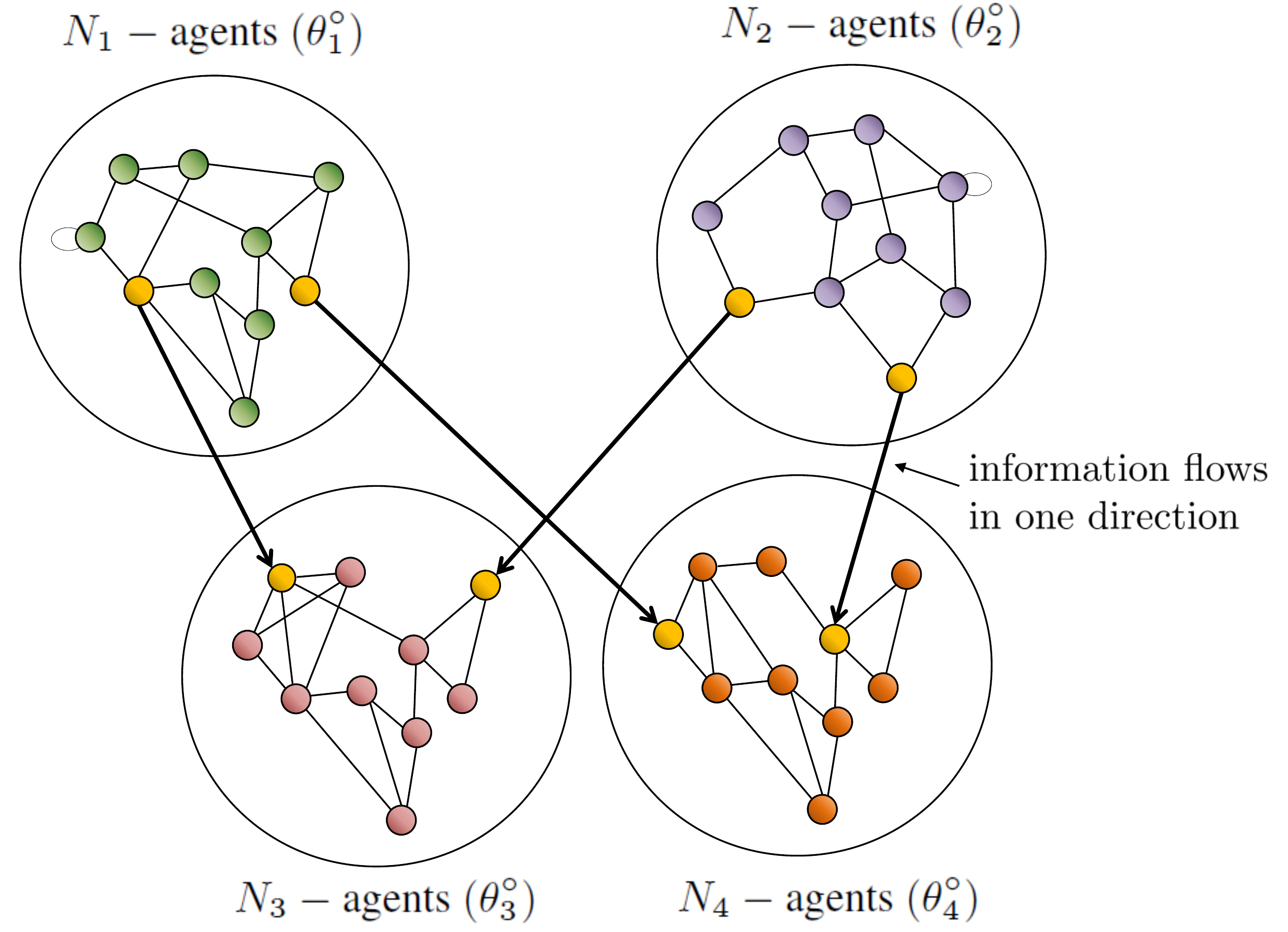} 
	\caption{ An example of a weakly connected network. The two sub-networks on top are $S-$type, while the two sub-networks in the bottom are $R-$type. Observe how links from $S-$networks to $R-$networks flow in one direction only, while all other links can be bi-directional.}
	\label{fig.WCFigure}
\end{figure}

We start the exposition by  reviewing the structure of weak graphs from \cite{ying2014information,icassp2,salamijournal} and by  introducing  the relevant notation. As explained in \cite{ying2014information}, a weakly-connected network consists of two types of sub-networks: ${S}$ (sending) sub-networks and ${R}$ (receiving) sub-networks. Each individual sub-network is a {\em connected} graph where any two agents are connected by a path. In addition, every sending sub-network is {\em strongly}-connected, meaning that at least one of its agents has a self-loop. The flow of information between $S$ and $R$ sub-networks is asymmetric, as it only happens in one direction from $S$ to $R$. Figure \ref{fig.WCFigure} shows one example of a weakly-connected network. The two top sub-networks are sending sub-networks and the two bottom sub-networks are receiving sub-networks. The weights on the connections from $S$ to $R$ networks are positive but can be arbitrarily small. Observe how links from $S-$subnetworks to $R-$subnetworks flow in one direction only, while all other links can be bi-directional. 

We index the strongly-connected sub-networks by $s=\{1,2,\cdots,S\}$, and the receiving sub-networks by $r=\{S+1,\dots,S+R\}$. Each sub-network $s$ has $N_s$ agents, and the total number of agents in the $S$ sub-networks is denoted by $N_{gS}$. Similarly, each sub-network $r$ has $N_{r}$ agents, and the total number of agents in the $R$ sub-networks is denoted by $N_{gR}$. We let $N$ denote the total number of agents across all sub-networks, i.e., $N=N_{gS}+N_{gR}$, and use $\mathcal{N}=\{1,2,\cdots,N\}$ to refer to the indexes of all agents. We assign a pair of non-negative weights, $\{a_{k\ell},a_{\ell k}\}$, to the edge connecting any two agents $k$ and $\ell$. The scalar $a_{\ell k}$ represents the weight with which agent $k$ scales data arriving from agent $\ell$ and, similarly, for $a_{k\ell}$. We let $\mathcal{N}_k$ denote the neighborhood of agent $k$, which consists of all agents connected to $k$. Each agent $k$ scales data arriving from its neighbors in a convex manner, i.e., the weights satisfy:
\be
a_{\ell k}\geq 0, \quad \sum_{\ell \in \mathcal{N}_k}a_{\ell k}=1, \quad a_{\ell k}=0 \text{ if } \ell\notin\mathcal{N}_k \label{convexCond}
\ee
Following \cite{ying2014information,salamijournal}, and without loss in generality, we assume that the agents are numbered such that the indexes of $\mathcal{N}$ represent first the agents from the $S$ sub-networks, followed by those from the $R$ sub-networks. In this way, if we collect the $\{a_{\ell k}\}$ into a large $N \times N$ combination matrix $A$, then this matrix will have an upper block-triangular structure of the following form:

\begin{footnotesize}
	\begin{eqnarray}
	\nonumber
	\begin{array}{cc}
	\hspace{-0.1cm}\overbrace{\rule{25mm}{0mm}}^{\mathrm{Sub-networks:} 1,2,\ldots, S} &\; \overbrace{\rule{55mm}{0mm}}^{\mathrm{Sub-networks:} S+1, S+2, \ldots,S+R}
	\end{array}
	\\ \nonumber
	\hspace{-0.1cm}
	\left[
	\begin{array}{cccc|cccc}
	A_{1} 	&	0 	&\hdots	& 0 		&	A_{1,S+1}\hspace{-0.3cm} & A_{1,S+2}		 &\hdots		 &A_{1,S+R}\\	
	0          	& A_{2}   		&\hdots	& 0 		& 	A_{2,S+1} \hspace{-0.3cm}		& A_{2,S+2}	 	 &\hdots		 &A_{2,S+R}	\\
	\vdots 	& \vdots		&\ddots	&\vdots	& 	\vdots 			& \vdots			 &\ddots		 &\vdots	 \\
	0          	& 	0  		&\hdots	& A_{S} 	& 	A_{S,S+1} \hspace{-0.3cm}		& A_{S,S+2}	 	 &\hdots		 &A_{S,S+R}	\\
	\hline
	0          	& 	0  		&\hdots	& 0 		& 	A_{S+1} \hspace{-0.3cm}			& A_{S+1,S+2}	 	 &\hdots		 &A_{S+1,S+R}	\\
	0          	& 	0  		&\hdots	& 0 		& 	A_{S+2,S+1}	\hspace{-0.3cm}	 		& A_{S+2}	 	 	 &\hdots		 &A_{S+2,S+R}	\\
	\vdots 	& \vdots		&\ddots	&\vdots	& 	\vdots 			& \vdots			 &\ddots		 &\vdots	 \\
	0&0  		&\hdots	& 0 		& 	A_{S+R,S+1} \hspace{-0.3cm}       	& 	A_{S+R,S+2}			 	 &\hdots		 &A_{S+R}	 \\
	\end{array}
	\right]
	\end{eqnarray}\end{footnotesize}
\be
\label{matrixA}
\ee 

\noindent The matrices $\{A_1,\cdots,A_S\}$ on the upper left corner are left-stochastic primitive matrices corresponding to the $S$ strongly-connected sub-networks. Likewise, the matrices $\{A_{S+1},\cdots,A_{S+R}\}$ in the lower right-most block correspond to the internal weights of the $R$ sub-networks. We denote the block structure of $A$ in (\ref{matrixA}) by:
\be
A \define \ba{ccc}T_{SS}&\vline&T_{SR}\\\hline 0&\vline &T_{RR}\ea \label{AStruct}
\ee 

\textit{Notation:} We use lowercase letters to denote vectors, uppercase letters for matrices, plain letters for deterministic variables, and boldface for random variables. We also use $(.)\tran$ for transposition, $(.)^{-1}$ for matrix inversion, and $\preceq$ and $\succeq$ for vector element-wise comparisons.

\section{Diffusion Social learning}
In order to characterize the set of attainable states, and to design mechanisms for belief control over weak graphs, we need to summarize first the main finding from \cite{salamijournal}. The work in that reference revealed the limiting states that are reached by receiving agents over weak-graphs. An expression was derived for these states. Once we review that expression, we will then examine its implications closely. In particular, we will conclude from it that not all states are attainable and that receiving sub-networks have an inherent resistance mechanism. We characterize this mechanism analytically. We then show how sending sub-networks can exploit this information to control the beliefs of receiving agents and to sow discord among the agents. 

Thus, following \cite{salamijournal}, we assume that each sub-network is observing data that arise from a true state value, denoted generically by $\theta^\circ$, which may differ from one sub-network to another. We denote by $\Theta$ the set of all possible states, by $\theta^\circ_s$ the true state of sending sub-network $s$ and by $\theta_r^\circ$ the true state of receiving sub-network $r$, where both $\theta^\circ_s$ and $\theta^\circ_r$ are in $\Theta$. At each time $i$, each agent $k$ will possess a belief $\bm\mu_{k,i}(\theta)$, which represents a probability distribution over $\theta \in \Theta$. Agent $k$ continuously updates its belief according to two information sources:
\begin{enumerate}
	\item The first source consists of observational signals $\{\bm\xi_{k,i}\}$ streaming in locally at agent $k$. These signals are generated according to some known likelihood function parametrized by the true state of agent $k$. We denote the likelihood function by $L_k(.|\theta^\circ_r)$ if agent $k$ belongs to receiving sub-network $r$ or $L_k(.|\theta^\circ_s)$ if agent $k$ belongs to sending sub-network $s$.
	\item The second source consists of information received from the neighbors of agent $k$, denoted by ${\cal N}_k$. Agent $k$ and its neighbors are connected by edges and they continuously communicate and share their opinions.
\end{enumerate} 

Using these two pieces of information, each agent $k$ then updates its belief according to the following diffusion social learning rule \cite{zhao2012learning}:
	\be
	\left\{
	\begin{aligned}
		\bm\psi_{k,i}(\theta) & = \frac{\bmu_{k,i-1}(\theta)L_k(\bm\xi_{k,i}|\theta)}{\sum_{\theta'\in\Theta}\bmu_{k,i-1}
			(\theta')L_k(\bm\xi_{k,i}|\theta')} \\ 
		\bmu_{k,i}(\theta) & = \sum_{\ell\in\mathcal{N}_k}a_{\ell k}\,\bm\psi_{\ell,i}(\theta)
		\label{eqn:diffusion} 
	\end{aligned}
	\right.
	\ee
	In the first step of (\ref{eqn:diffusion}), agent $k$ updates its belief, $\bm\mu_{k,i-1}(\theta)$, based on its observed private signal $\bm\xi_{k,i}$ by means of the Bayesian rule and obtains an intermediate belief $\bm\psi_{k,i}(\theta)$. In the second step, agent $k$ learns from its social neighbors through cooperation. 
	
	A consensus-based strategy can also be employed in lieu of (\ref{eqn:diffusion}), as was done in the insightful works \cite{jadbabaie2012non,mouraa}, although the latter reference focuses mainly on the problem of pure averaging and not on social learning and requires the existence of certain anchor nodes. In this work, we assume all agents are homogeneous and focus on the diffusion strategy (\ref{eqn:diffusion}) due to its enhanced performance and wider stability range, as already proved in \cite{zhao2012learning} and further explained in the treatments \cite{Sayed,sayed2014adaptive}. Other models for social learning can be found in \cite{Acemoglu,Yildiz,krishna1,misinformation,latiha,optiJad,modi1}.

When agents of sending sub-networks follow this model, they can learn their own true states. Specifically, it was shown in \cite{zhao2012learning,salamijournal} that
\begin{align}
\lim_{i\rightarrow\infty} \bm \mu_{k,i}(\theta^\circ_s) \aseq 1
\label{truthLearning}
\end{align}
for any agent $k$ that belongs to sending sub-network $s$. Result (\ref{truthLearning}) means that the probability measure concentrates at location $\theta_s^\circ$, while all other possibilities in $\Theta$ have zero probability. On the other hand, agents of receiving sub-networks will not be able to find their true states. Instead, their beliefs will converge to a fixed distribution defined over the true states of the {\em sending} sub-networks as follows \cite{salamijournal}. First, let
\be
\bmu_{i}^s (\theta)\define
\ba {c}
\bmu_{k_s(1),i} (\theta) \\
\bmu_{k_s(2),i} (\theta) \\
\vdots \\
\bmu_{k_s(N_s),i}(\theta) 
\ea
\ee 
collect all beliefs from agents that belong to sub-network $s$, where the notation $k_s(n)$ denotes the index of the $n${-th} agent within sub-network $s$, i.e.,
\be 
k_s(n)=\sum\limits_{v=1}^{s-1}N_v+n
\ee
and $n\in \{1,2,\cdots,N_s\}$.
Likewise, let
\be
\bmu_{i}^r (\theta)\define
\ba {c}
\bmu_{k_r(1),i} (\theta) \\
\bmu_{k_r(2),i} (\theta) \\
\vdots \\
\bmu_{k_r(N_{r}),i}(\theta) 
\ea
\ee 
collect all beliefs from agents that belong to sub-network $r$, where the notation $k_r(n)$ denotes the index of the $n$-th agent within sub-network $r$, i.e.,
\be
k_r(n)=N_{gS}+\sum\limits_{v=S+1}^{r-1}N_{v}+n
\label{Index}
\ee
and $n\in \{1,2,\cdots,N_r\}$. Furthermore, let
\be
\bmu_{\mathcal{S},i} (\theta)\define
\ba {c}
\bmu_{i}^1 (\theta) \\
\vdots \\
\bmu_{i}^S(\theta) 
\ea
\label{muS}
\ee
collect all beliefs from all $S-$type sub-networks. Likewise, let
\be
\bmu_{\mathcal{R},i} (\theta)\define
\ba {c}
\bmu_{i}^{S+1} (\theta) \\
\vdots \\
\bmu_{i}^{S+R}(\theta) 
\ea
\label{muR}
\ee
collect the beliefs from all $R-$type sub-networks. Note that these belief vectors are evaluated at a specific $\theta \in \Theta$. Then, the main result in \cite{icassp2,salamijournal} shows that, under some reasonable technical assumptions, it holds that
\be	
\lim_{i \to \infty} \bmu_{\mathcal{R},i} (\theta) = W\tran \left (\lim_{i \to \infty}  \bmu_{\mathcal{S},i} (\theta)\right) 
\label{result1}
\ee
where $W$ is the $N_{gS} \times N_{gR}$ matrix given by:
\bq
W &\define & T_{SR}(I-T_{RR})^{-1} \label{defW1}
\eq
and $I$ is the identity matrix of size $N_{gR}$. The matrix $W$ has non-negative entries and the sum of the entries in each of its columns is equal to one \cite{ying2014information}. Expression (\ref{result1}) shows how the beliefs of the sending sub-networks determine the limiting beliefs of the receiving sub-networks through the matrix $W$. We can expand (\ref{result1}) to reveal the influence of the sending networks more explicitly as follows. 

Let $w_{k}\tran$ denote the row in $W\tran$ that corresponds to receiving agent $k$ and partition it into sub-vectors as follows{\footnote{The index of the row in $W\tran$ that corresponds to agent $k$ is $k-N_{gS}$.}:
\begin{align}
w_k\tran=\ba{ccccccc}w_{k,N_1}\tran&\vline&w_{k,N_2}\tran&\vline&\ldots&\vline&w_{k,N_S}\tran\ea
\end{align}
where the $\{N_1,N_2,\ldots,N_S\}$ are the number of agents in each sub-network $s\in\{1,2,\ldots,S\}$.
Then, according to (\ref{result1}), we have
\begin{align}
\lim_{i \to \infty}  \bmu_{k,i} (\theta) =\ba{ccccc}w_{k,N_1}\tran&w_{k,N_2}\tran&\ldots&w_{k,N_S}\tran\ea  \left (\lim_{i \to \infty}  \bmu_{\mathcal{S},i} (\theta)\right) \label{exaplanation}
\end{align}
Note that this relation is for a specific $\theta \in \Theta$. Let us focus on the case when $\theta=\theta_s^{\circ}$, assuming it is the true state parameter of the $s$-th sending network only. We know from \cite{zhao2012learning} and (\ref{truthLearning}) that each agent in the sending sub-network $s$ will learn its true state $\theta_s^\circ$. Therefore, from (\ref{muS}),
\begin{align}
\lim_{i \to \infty}  \bmu_{\mathcal{S},i} (\theta^\circ_s)= \ba{c}
\boldsymbol{0}_{N_1}\\
\boldsymbol{0}_{N_2}\\
\vdots\\
\one_{N_s}\\
\vdots\\
\boldsymbol{0}_{N_S}
\ea \label{finals}
 \end{align}
where $\one_{N_s}$ denotes a column vector of length $N_s$ whose elements are all one. Similarly, $\boldsymbol{0}_{Ns}$ denotes a column vector of length $N_s$ whose elements are all zero. Combining (\ref{exaplanation}) and (\ref{finals}) we get
\begin{align}
\lim_{i \to \infty} \bm\mu_{k,i}(\theta^\circ_s)=w_{k,N_s}\tran\one_{N_s}
\end{align}
    
\noindent This means that the likelihood of state $\theta_s^\circ$ at the receiving agent $k$ is equal to the sum of the entries of the weight vector, $w_{k,N_s}$, corresponding to sub-network $s$. More generally, for any other state parameter $\theta\in\Theta$, its likelihood is given from (\ref{result1}) by 
\begin{align}
	\lim_{i \to \infty} \bm\mu_{k,i}(\theta)= \sum_{s=1}^{S}w_{k,N_s}\tran e_{\theta,\theta^\circ_s}
	\label{FinalDistribution}
\end{align}
where
		\be
		e_{\theta,\theta^\circ_s}
		\define \left\{
		\begin{aligned}
			\one_{N_s},&\quad {\rm if}\quad \theta = \theta_s^\circ\\
			\boldsymbol{0}_{N_s},&\quad {\rm otherwise}
		\end{aligned}
		\right. 
		\label{eqn:delta2}
		\ee

	\noindent Result (\ref{FinalDistribution}) means that the belief of receiving agent $k$ will converge to a distribution defined over the true states of the sending sub-networks, which we collect into the set:
	\begin{align}
	\Theta^{\bullet}\define\{\theta_1^{\circ},\theta_2^{\circ},\hdots,\theta_S^{\circ}\}
	\end{align} 
   
  Expression (\ref{result1}) shows how the limiting distributions of the sending sub-networks determine the limiting distributions of the receiving sub-networks through the matrix $W\tran$. In other words, it indicates how influential agents (from within the sending sub-networks) can control the steady-state beliefs of receiving agents. Two critical questions arise at this stage: (a) first, how much freedom do influential agents have in controlling the beliefs of the receiving agents? That is, can receiving agents be driven to arbitrary beliefs or does the network structure limit the scope of control by the influential agents? and (b) second, even if there is a limit to what influential agents can accomplish, how can they ensure that receiving agents will end up with particular beliefs?

 Questions (a) and (b) raise interesting possibilities about belief (or what we will sometimes refer to as ``mind'') control. In the next sections, we will address these questions and we will end up with the conditions that allow influential agents to drive other agents to endorse particular beliefs regardless of their local observations (or ``convictions'').

 \section{Belief Control Mechanism}
 Observe from expression (\ref{FinalDistribution}) that the limiting beliefs of receiving agents depend on the columns of $W=T_{SR}\left(I-T_{RR}\right)^{-1}$. Note also that the entries of $W$ are determined by the internal combination weights within the receiving networks (i.e., $T_{RR}$), and the combination weights from the $S$ to the $R$ sub-networks (i.e., $T_{SR}$). The question we would like to examine now is that given a set of desired beliefs for the receiving agents, is this set always attainable? Or does the internal structure of the receiving sub-networks impose limitations on where their beliefs can be driven to? To answer this useful question, we consider the following problem setting. Let $q_k(\theta)$ denote some desired limiting distribution for receiving agent $k$ (i.e., $q_k(\theta)$ denotes what we desire the limiting distribution $\bm{\mu}_{k,i}(\theta)$ in (\ref{FinalDistribution}) to become as $i\rightarrow\infty$). We would like to examine whether it is possible to force agent $k$ to converge to {\emph{any}} $q_k(\theta)$, i.e., whether it is possible to find a matrix $T_{SR}$ so that the belief of receiving agent $k$ converges to this specific $q_k(\theta)$.

\subsection{Motivation}
In this first approach, we are interested in designing $T_{SR}$ while $T_{RR}$ is assumed fixed and known. This scenario allows us to understand in what ways the internal structure of the receiving networks limits the effect of external influence by the sending sub-networks. This approach also allows us to examine the range of belief control over the receiving sub-networks (i.e., how much freedom the sending sub-networks have in selecting these beliefs). Note that the entries of $T_{SR}$ correspond to weights by which the receiving agents scale information from the sending sub-networks. These weights are set by the receiving agents and, therefore, are not under the direct control of the sending sub-networks. As such, it is fair to question whether it is useful to pursue a design procedure for selecting $T_{SR}$ since its entries are not under the direct control of the designer or the sending sub-networks. The useful point to note here, however, is that the entries of $T_{SR}$, although set by the receiving agents, can still be interpreted as a measure of the level of trust that receiving agents have in the sending agents they are connected to. The higher this level of confidence is between two agents, the larger the value of the scaling weight on the link connecting them. In many applications, these levels of confidence (and, therefore, the resulting scaling weights) can be influenced by external campaigns (e.g., through advertisement or by way of reputation). In this way, we can interpret the problem of designing $T_{SR}$ as a way to guide the campaign that influences receiving agents to set their scaling weights to desirable values. The argument will show that by influencing and knowing $T_{SR}$, sending agents end up controlling the beliefs of receiving agents in desirable ways. For the analysis in the sequel, note that by fixing $T_{RR}$ and designing $T_{SR}$, we are in effect fixing the sum of each column of $T_{SR}$ and, accordingly, fixing the overall external influence on each receiving agent. In this way, the problem of designing $T_{SR}$ amounts to deciding on how much influence each individual sub-network should have in driving the beliefs of the receiving sub-networks. 

\subsection{Conditions for Attainable Beliefs}
 Given these considerations, let us now show how to design $T_{SR}$ to attain  certain beliefs. As is already evident from (\ref{FinalDistribution}), the desired belief $q_k(\theta)$ at any agent $k$ needs to be a probability distribution defined over the true states of all {\em{sending}} sub-networks, $\Theta^{\bullet}=\{\theta_1^{\circ},\theta_2^{\circ},\hdots,\theta_S^{\circ}\}$. We assume, without loss of generality, that the true states of the sending sub-networks are distinct, so that $|\Theta^{\bullet}|=S$. If two or more sending sub-networks have the same true state, we can merge them together and treat them as corresponding to one sending sub-network; although this enlarged component is not necessarily connected, it nevertheless consists of strongly-connected elements and the same arguments and conclusions will apply. 
 
 We collect the desired limiting beliefs for all receiving agents into the vector:
 \begin{align}
 q_{\mathcal{R}} (\theta)\define
 \ba{c}
 q_{N_{gS}+1} (\theta) \\
 q_{N_{gS}+2} (\theta) \\
 \vdots\\
 q_{N}(\theta) 
 \ea
 \label{muqR}
\end{align}
 which has length $N_{gR}$. Then, from (\ref{result1}), we must have:
 \begin{align}
  q_{\mathcal{R}}\tran(\theta)= \left (\lim_{i \to \infty}  \bmu_{\mathcal{S},i} (\theta)\right)\tran W \label{exp1}
 \end{align}
Evaluating this expression at the successive states $\{\theta_1^\circ, \theta_2^\circ,\ldots,\theta_S^\circ\}$, we get 
  \begin{align}
 \underbrace{\ba{c}
 q_{\mathcal{R}}\tran(\theta_1^{\circ})\\
 q_{\mathcal{R}}\tran(\theta_2^{\circ})\\
 \vdots\\
 q_{\mathcal{R}}\tran(\theta_S^{\circ})
 \ea}_{\define Q_{S\times N_{gR}}}
 &=\underbrace{\ba {c c c c}
 \one_{N_1}\tran&\boldsymbol{0}&\hdots&\boldsymbol{0}\\
 \boldsymbol{0}&\one_{N_2}\tran&\hdots&\boldsymbol{0}\\
 \vdots&\vdots&\ddots&\vdots\\
 \boldsymbol{0}&\boldsymbol{0}&\hdots&\one_{N_{S}}\tran
 \ea }_{\define E_{S\times N_{gS}}}W 
  \label{defE}
  \end{align}
 where $Q$ is the $S \times N_{gR}$ matrix that collects the desired beliefs for all receiving agents. Using (\ref{defW1}), we rewrite (\ref{defE}) more compactly in matrix form as:
\begin{align}
E\;T_{SR}=Q\left(I-T_{RR}\right) \label{condition}
\end{align}
Therefore, given $Q$ and $T_{RR}$, the design problem becomes one of finding a matrix $T_{SR}$ that satisfies (\ref{condition}) subject to the following constraints: 
\begin{align}
 \one\tran T_{SR}+\one\tran T_{RR}&=\one\tran \label{cond1}\\
 T_{SR}&\succcurlyeq 0 \label{cond2}\\
 t_{SR,k}(j)&=0 \text{, if receiving agent $k$ is not}\nn\\
 \phantom{T_{SR}(j,k)=0}&\phantom{,,,,,}\text{connected to sending agent $j$ } \label{cond3}
 \end{align}
 The first condition (\ref{cond1}) is because the entries on each column of $A$ defined in (\ref{AStruct}) add up to one. The second condition (\ref{cond2}) ensures that each element of $T_{SR}$ is a non-negative combination weight. The third condition (\ref{cond3}) takes into account the network structure, where $t_{SR,k}$ represents the column of $T_{SR}$ that corresponds to receiving agent $k$, and $t_{SR,k}(j)$ represents the $j^{th}$ entry of this column (which corresponds to sending agent $j$--see Fig. \ref{imageTSR}). In other words, if receiving agent $k$ is not connected to sending agent $j$, the corresponding entry in $T_{SR}$ should be zero.
 \begin{figure}[h]
 	\centering
 	\includegraphics[scale=0.5]{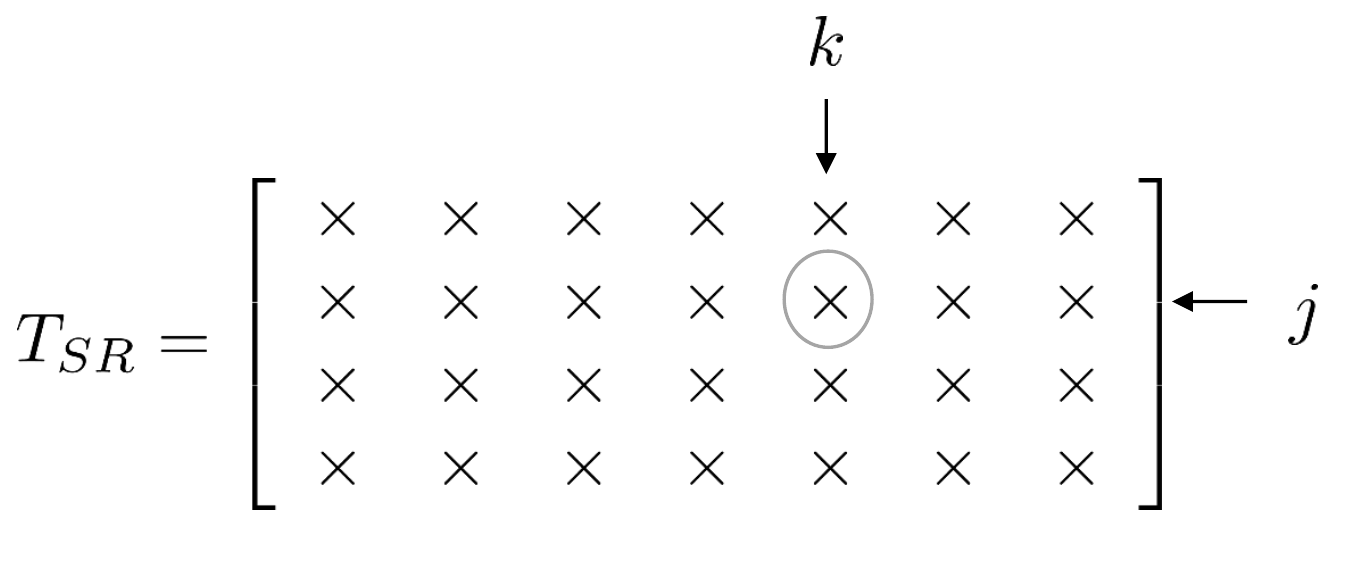}
 	\caption{An illustration of the $k-$th column of $T_{SR}$ and the $j-$th entry on that column.}
 	\label{imageTSR}
 \end{figure}  
 
 It is useful to note that condition (\ref{cond1}) is actually unnecessary and can be removed. This is because if we can find $T_{SR}$ that satisfies (\ref{condition}), then condition (\ref{cond1}) will be automatically satisfied. To see this, we first sum the elements of the columns on the left-hand side of (\ref{condition}) and observe that
 \begin{align}
 \one\tran_{S}E T_{SR}=\one\tran T_{SR} \label{rell1}
 \end{align}
 We then sum the elements of the columns on the right-hand side of (\ref{condition}) to get
 \begin{align}
 \one_{S}\tran\left(Q-QT_{RR}\right)=\one\tran-\one\tran T_{RR} \label {rell2}
 \end{align}
 This is because $\one_S\tran Q=\one_{N_{gR}}\tran$ since the entries on each column of $Q$ add up to one. Thus, equating (\ref{rell1}) and (\ref{rell2}), we find that (\ref{cond1}) must hold. The problem we are attempting to solve is then equivalent to finding $T_{SR}$ that satisfies (\ref{condition}) subject to
 \begin{align}
 T_{SR}&\succcurlyeq 0 \label{cond22} \\
 t_{SR,k}(j)&=0 \text{, if receiving agent $k$ is not}\nn\\
 \phantom{T_{SR}(j,k)=0}&\phantom{,,,,,}\text{connected to sending agent $j$ } \label{cond33}
 \end{align}
To find $T_{SR}$ that satisfies (\ref{condition}) under the constraints (\ref{cond22})-(\ref{cond33}), we can solve separately for each column of $T_{SR}$. Let $t_{RR,k}$ and $q_k$, respectively, denote the columns of $T_{RR}$ and $Q$ that correspond to receiving agent $k$. Then, relations (\ref{condition}) and (\ref{cond22})--(\ref{cond33}) imply that column $t_{SR,k}$ must satisfy:
\begin{align}
E t_{SR,k}=q_k-Qt_{RR,k} \label{relation}
\end{align}
subject to
\begin{align}
t_{SR,k}&\succcurlyeq 0 \label{cond2vec} \\
t_{SR,k}(j)&=0\text{, if receiving agent $k$ is not}\nn\\
\phantom{T_{SR}(j,k)=0}&\phantom{,,,,,}\text{connected to sending agent $j$ } \label{cond3vec}
\end{align}
The problem is then equivalent to finding $t_{SR,k}$ for each receiving agent $k$ such that $t_{SR,k}$ satisfies (\ref{relation})-(\ref{cond3vec}). For $Q$ to be attainable (i.e., for the beliefs of all receiving agents to converge to the desired beliefs), finding such $t_{SR,k}$ should be possible for each receiving agent $k$. However, 
 finding $t_{SR,k}$ that satisfies (\ref{relation}) under the constraints (\ref{cond2vec})-(\ref{cond3vec}) may not be always possible. The desired belief matrix $Q$ will need to satisfy certain conditions so that it is not possible to drive the receiving agents to any belief matrix $Q$. Before stating these conditions, we introduce two auxiliary matrices. We define first the following difference matrix, which appears on the right-hand side of (\ref{condition}) --- this matrix is known:
 \be
 V\define Q(I-T_{RR})
 \ee
 Note that $V$ has dimensions $S\times N_{gR}$. The $k-$th column of $V$, which we denote by $v_k$ appears on the right-hand side of (\ref{relation}), i.e.,
 \begin{align}
 v_{k}=q_k-Qt_{RR,k}
 \end{align}
 The $(s,k)-$th entry of $V$ is then:
 \begin{align}
v_{k}(s)=q_k(\theta^\circ_s)-\sum_{\ell=1}^{N_{gR}}t_{RR,k}(\ell)q_{
	N_{gS}+\ell}(\theta^\circ_s)
 \end{align}
 
 Each $(s,k)-$th entry of $V$  represents the difference between the desired limiting belief at $\theta^\circ_s$ of receiving agent $k$ and a weighted combination of the desired limiting beliefs of its neighboring receiving agents. We remark that this sum includes agent $k$ if $t_{RR,k}(k)$ is not zero. Similarly, it includes any receiving agent $\ell$ if $t_{RR,k}(\ell)$ is not zero. In this way, the sum runs only over the neighbors of agent $k$, because any agent $\ell$ that is not a neighbor of agent $k$ has its corresponding entry in $t_{RR,k}$ as zero.
 	
 Let $C$ denote an $S\times N_{gR}$ binary matrix, with as many rows as the number of sending sub-networks  and as many columns as the number of receiving agents. The matrix $C$ is an indicator matrix that specifies whether a receiving agent is connected or not to a sending sub-network. The $(s,k)-$th entry of $C$ is one if receiving agent $k$ is connected to sending sub-network $s$; otherwise, it is zero. We are now ready to state when a given set of desired beliefs is attainable.
 
  \begin{theorem}({\bf Attainable Beliefs}) A given belief matrix $Q$ is attainable if, and only if, the entries of $V$ will be zero wherever the entries of $C$ are zero, and the entries of $V$ will be positive wherever the entries of $C$ are  one. 
  	 \qd 
   \end{theorem}
  
 Before proving theorem 1, we first clarify its statement. For $Q$ to be achievable, the matrices $V$ and $C$ must have the same structure with the unit entries of $C$ translated into positive entries in $V$. This theorem reveals two possible cases for each receiving agent $k$ and gives, for each case, the condition required for the desired beliefs to be attainable. 
 
 In the first case, receiving agent $k$ is not connected to any agent of sending sub-network $s$ (the  $(s,k)-$th entry of $C$ is zero). Then, according to Theorem 1, receiving agent $k$ achieves its desired limiting belief $q_k(\theta^\circ_s)$ if, and only if,
 \begin{align}
  v_{k}(s)=q_k(\theta^\circ_s)-\sum_{\ell=1}^{N_{gR}}t_{RR,k}(\ell)q_{
  	N_{gS}+\ell}(\theta^\circ_s)=0 \label{noCon}
  \end{align}
 That is, the cumulative influence from the agent's neighbors must match the desired limiting belief. 
 
 In the second case, receiving agent $k$ is connected to at least one agent of sending sub-network $s$ (the  $(s,k)-$th entry of $C$ is one). Now, according to Theorem 1 again, receiving agent $k$ achieves its desired limiting belief $q_k(\theta^\circ_s)$ if, and only if, 
 \begin{align}
v_{k}(s)=q_k(\theta^\circ_s)-\sum_{\ell=1}^{N_{gR}}t_{RR,k}(\ell)q_{
	N_{gS}+\ell}(\theta^\circ_s)> 0 \label{wtfCon}
 \end{align} \begin{proof}[Proof of Theorem 1] We start by first proving that if $Q$ is attainable, then $V$ and $C$ have the same structure. If $Q$ is attainable, then there exists $t_{SR,k}$ for each receiving agent $k$ that satisfies (\ref{relation})-(\ref{cond3vec}).  Using the definition of $E$ in (\ref{defE}), the $s-$th row on the left-hand side of (\ref{relation}) is: 
\begin{align}
\sum_{j\in\mathcal{I}_s}t_{SR,k}(j) \label{block}
\end{align}
where $\mathcal{I}_s$ represents the set of indexes of sending agents that belong to sending sub-network $s$. Expression (\ref{block}) represents the sum of the elements of the block of $t_{SR,k}$ that correspond to sending sub-network $s$. Therefore, if $Q$ is attainable, then the $s-$th row of (\ref{relation}) satisfies the following relation:
\begin{align}
\sum_{j\in\mathcal{I}_s}t_{SR,k}(j)=v_k(s) \label{relationS}
\end{align}
From this relation, we see that if agent $k$ is not connected to any agent in sub-network $s$, then $\sum_{j\in\mathcal{I}_s}t_{SR,k}(j)=0$ which implies that $v_k(s)$ is zero. On the other hand, if agent $k$ is connected to sub-network $s$, then $\sum_{j\in\mathcal{I}_s}t_{SR,k}(j)>0$ which implies that $v_k(s)>0$. In other words, $C$ and $V$ have the same structure.

Conversely, if $C$ and $V$ have the same structure, then it is possible to find $t_{SR,k}$ for each receiving agent $k$ that satisfies (\ref{relation})-(\ref{cond3vec}). In particular, if agent $k$ is not connected to sub-network $s$, then the ($s,k$)-th entry of $C$ is zero. Since $C$ and $V$ have the same structure, then $v_k(s)=0$ . By setting to zero the entries of $t_{SR,k}$ that correspond to sending sub-network $s$, relation (\ref{relationS}) is satisfied. On the other hand, if agent $k$ is connected to sub-network $s$ (connected to at least one agent in sub-network $s$), then the ($s,k$)-th entry of $C$ is one.  Since $C$ and $V$ have the same structure, we get $v_k(s)>0$. Therefore, since the entries of $t_{SR,k}$ must be non-negative, we first set to zero the entries of $t_{SR,k}$ that correspond to agents of sub-network $s$ that are not connected to agent $k$ and the remaining entries can be set to non-negative values such that relation (\ref{relationS}) is satisfied. That is, if $C$ and $V$ have the same structure, then $Q$ is attainable. 
\end{proof}

We next move to characterize the set of solutions, i.e., how we can design $t_{SR,k}$ assuming the conditions on $V$ are met.
\subsection{Characterizing the Set of Possible Solutions}
In the sequel, we assume that the conditions on $V$ from Theorem 1 are satisfied. That is, if receiving agent $k$ is not connected to sub-network $s$, then $v_k(s)=0$. Otherwise, $v_k(s)> 0$. The desired beliefs are then attainable. This means that for each receiving agent $k$, we can find $t_{SR,k}$ that satisfies (\ref{relation})-(\ref{cond3vec}). Many solutions may exist. In this section, we characterize the set of possible solutions.   

First of all, to meet (\ref{cond33}), we set the required entries of $t_{SR,k}$ to zero. We then remove the corresponding columns of $E$, and label the reduced $E$ by $E_k$. Similarly, we remove the zero elements of $t_{SR,k}$ and label the reduced $t_{SR,k}$ by $t'_{SR,k}$. On the other hand, if agent $k$ is not connected to some sub-network $s$, then the corresponding row in $E$ will be removed and $E_k$ will have fewer number of rows, denoted by $S'$.  Without loss of generality, we assume agent $k$ is connected to the first $S'$ sending sub-networks.  We denote by $N_s^k$ the number of agents of sending sub-network $s$ that are connected to receiving agent $k$ and by $N_{gS}^k$ the total number of all sending agents connected to agent $k$. The matrix $E_k$ will then have the form (this matrix is obtained from $E$ by removing rows and columns with zero entries; the resulting dimensions are now denoted by $S'$ and $N_{gS}^k$):
\begin{align}
E_k=\ba {c c c c}
\one_{N_1^k}\tran&\boldsymbol{0}&\hdots&\boldsymbol{0}\\
\boldsymbol{0}&\one_{N_2^k}\tran&\hdots&\boldsymbol{0}\\
\vdots& \vdots&\ddots &\vdots \\
\boldsymbol{0}&\boldsymbol{0}&\hdots&\one_{N_{S'}^k}\tran
\ea _{S'\times N_{gS}^k}\label{EK} 
\end{align}
Note that if receiving agent $k$ is connected to all sending sub-networks, then $E$ and $E_k$ will have the same number of rows, $S'=S$. In the case where agent $k$ is not connected to some sub-network $s$, condition (\ref{noCon}) should be satisfied, and the corresponding row in $q_k-Qt_{RR,k}$ should be removed to obtain the reduced vector $q'_k-Q't_{RR,k}$. We are therefore reduced to determining $t_{SR,k}'$ by solving a system of equations of the form:
\begin{align}
E_k t'_{SR,k}=q'_k-Q't_{RR,k} \label{relation3}
\end{align}
subject to
\begin{align}
t'_{SR,k}&\succcurlyeq 0 \label{finalCond} 
\end{align} 
{We can still have some of the entries of the solution $t'_{SR,k}$ turn out to be zero}. Now note that the number of rows of $E_k$ is $S'$ (number of sending sub-networks connected to $k$), which is always smaller than or equal to $N_{gS}^k$. Moreover, the rows of $E_k$ are linearly independent and thus $E_k$ is a right-invertible matrix. Its right-inverse is given by \cite{Laub}:
\begin{align}
E^{R}_k=E_k\tran\left(E_kE_k\tran\right)^{-1}
\end{align}
Therefore, if we ignore condition (\ref{finalCond}) for now, then equation (\ref{relation3}) has an infinite number of solutions parametrized by the expression \cite{Laub}:
 \begin{align}
   t'_{SR,k}&=E_k^{R}\left(q'_k-Q't_{RR,k}\right)+(I-E_k^R E_k)y  \label{expS}
 \end{align}
 where $y$ is an arbitrary vector of length $N_{gS}^k$. We still need to satisfy condition (\ref{finalCond}). Let 
    \begin{align}
    v'_k \define q'_k-Q't_{RR,k} \label{vk}
    \end{align}
    and note that
    \begin{align}
     E_k^Rv'_k=\ba {c}
     \frac{v'_k(1)}{N_1^k} \one_{N_1^k}\\
    \frac{v'_k(2)}{N_2^k} \one_{N_2^k}\\
    \vdots\\
    \frac{v'_k(S')}{N_{S'}^k} \one_{N_{S'}^k}
    \ea
    \end{align}
    where $v'_k(i)$ represents the $i^{th}$ entry of vector $v'_k$. Likewise,
    \begin{align}
    I-E_k\tran\left(E_kE_k\tran\right)^{-1}E_k=\;\;\;\;\;\;\;\;\;\;\;\;\;\;\;\;\;& \nn \\
   \text{diag}  \left\{  I_{N_1^k}-\frac{1}{N_1^k} \one_{N_1^k}\one_{N_1^k}\tran, I_{N_2^k}-\frac{1}{N_2^k} \one_{N_2^k}\one_{N_2^k}\tran,\cdots\right.& \nn \\
     \left.I_{N_{S'}^k}-\frac{1}{N_{S'}^k} \one_{N_{S'}^k}\one_{N_{S'}^k}\tran  \right\}
   \end{align}
     and if we partition $y$ into sub-vectors as
      \be
      y=\ba{c}
      y_{N_1^k}\\
      y_{N_2^k}\\
      \vdots\\
      y_{N_{S'}^k}
      \ea
      \ee
  then expression (\ref{expS}) becomes:
   \begin{align}
    t'_{SR,k}=\ba {c}
       \frac{v'_k(1)}{N_1^k} \one_{N_1^k}\\
       \frac{v'_k(2)}{N_2^k} \one_{N_2^k}\\
       \vdots\\
       \frac{v'_k(S')}{N_{S'}^k} \one_{N_{S'}^k}
       \ea
       + 
       \ba {c}
       \left(I_{N_1^k}-\frac{1}{N_1^k}\one_{N_1^k}\one_{N_1^k}\tran\right)y_{N_1^k}\\
       \left(I_{N_2^k}-\frac{1}{N_2^k} \one_{N_2^k}\one_{N_2^k}\tran\right)y_{N_2^k}\\
       \vdots\\
        \left(I_{N_{S'}^k}-\frac{1}{N_{S'}^k} \one_{N_{S'}^k}\one_{N_{S'}^k}\tran\right) y_{N_{S'}^k} 
     \ea \label{tkFinal}
    \end{align}
      This represents the general form of all possible solutions, but from these solutions we want only those which are nonnegative in order to satisfy condition (\ref{finalCond}). From (\ref{tkFinal}), the vector $t_{SR,k}'$ is partitioned into multiple blocks, where each block has the form:
      \begin{align} 
      \frac{v'_k(s)}{N_{s}^k} \one_{N_{s}^k}+
       \left(I_{N_{s}^k}-\frac{1}{N_{s}^k} \one_{N_{s}^k}\one_{N_{s}^k}\tran\right) y_{N_s^k} \label{blocktk}
      \end{align}
   We already have from the conditions of attainable beliefs (\ref{wtfCon}) that $v'_k(s)> 0$. Therefore, we can choose $y_{N_s^k}$ as zero or set it to arbitrary values as long as (\ref{blocktk}) stays non-negative. We also know that for the beliefs to be attainable, we cannot have $v'_k(s)<0$. Otherwise, no solution can be found. Indeed, if $v'_k(s)<0$, then to make (\ref{blocktk}) non-negative, we would need to select $y_{N_s^k}$ such that:  
      \begin{align} 
      \left(I_{N_{s}^k}-\frac{1}{N_{s}^k} \one_{N_{s}^k}\one_{N_{s}^k}\tran\right) y_{N_s^k}\succeq -\frac{v'_k(s)}{N_{s}^k}\one_{N_{s}^k} \label{imp}
      \end{align}
      However, there is no $y_{N_s^k}$ that satisfies this relation because if we sum the elements of the vector on the left-hand side of (\ref{imp}), we obtain:
      \begin{align} 
       \one_{N_s^k}\tran\left(I_{N_{s}^k}-\frac{1}{N_{s}^k} \one_{N_{s}^k}\one_{N_{s}^k}\tran\right) y_{N_s^k}=0
      \end{align}
     While if we sum the elements of the vector on the right-hand side of (\ref{imp}), we obtain:
     \begin{align}
     -\frac{v'_k(s)}{N_{s}^k}\one_{N_{s}^k}\tran\one_{N_{s}^k}=-v'_k(s) >0
     \end{align}
     This means that we cannot find $t'_{SR,k}$ such that 
      $t'_{SR,k} \succeq 0$ when any of the entries of $v'_k$ or $q'_k-Q't_{RR,k}$ is negative. 
      
      In summary, we have established the validity of the following statement.
      
      \begin{theorem} Assume receiving agent $k$ is connected to $N_s^k$ agents in sending sub-network $s$. If $v_k(s)>0$, then all possible choices for the weights from sending agents in network $s$ to receiving agent $k$ are parameterized as:
      \begin{align} 
      \frac{v'_k(s)}{N_{s}^k} \one_{N_{s}^k}+
      \left(I_{N_{s}^k}-\frac{1}{N_{s}^k} \one_{N_{s}^k}\one_{N_{s}^k}\tran\right) y_{N_s^k} \label{blocktkT}
      \end{align}
       where $y_{N_s^k}$ is an arbitrary vector of length $N_s^k$ chosen so that (\ref{blocktkT}) stays non-negative.
       \qd
      \end{theorem}

     \subsection{Enforcing Uniform Beliefs}
     In this section, we explore one special case of attainable beliefs, which is driving all receiving agents towards the {\emph same} belief. In this case, $Q$ is of the following form:
      \begin{align}
      Q=q \one_{N_{gR}}\tran
      \end{align}
       for some column $q$ that represents the desired limiting belief (the entries of $q$ are non-negative and add up to one). We verify that the conditions that ensure that uniform beliefs are attainable by all receiving agents. In this case, $v_k$ is of the following form:
       \begin{align}
       v_k=q_k-Qt_{RR,k}=(1-\one_{N_{gR}}\tran t_{RR,k})q \label{Uni}
       \end{align}
       and the ($s,k$)-th entry of $V$ is:
       \begin{align}
        v_k(s)=(1-\one_{N_{gR}}\tran t_{RR,k})q(\theta^\circ_s)  \label{Uni2}
       \end{align}
 Now we know that $1-\one_{N_{gR}}\tran t_{RR,k}>0$ when agent $k$ is connected to at least one agent from any sending sub-network, and that $1-\one_{N_{gR}}\tran t_{RR,k}=0$ when it is not connected to any sending sub-network. In the second case where $1-\one_{N_{gR}}\tran t_{RR,k}=0$, expression (\ref{Uni2}) implies that $v_k(s)=0$ for any $s$. 
 Therefore, in this case, we have agent $k$ not connected to any sending sub-network $s$ and $v_k(s)=0$ for any $s$, and condition (\ref{noCon}) is satisfied.
 In the first case where $1-\one_{N_{gR}}\tran t_{RR,k}>0$ (i.e., agent $k$ is connected to some sending sub-networks but not necessarily to all of them), expression (\ref{Uni2}) implies that $v_k(s)>0$ no matter whether agent $k$ is connected or not to sending sub-network $s$. However, when agent $k$ is not connected to sending sub-network $s$, condition (\ref{noCon}) requires that $v_k(s)=0$ for agent $k$ to achieve its desired belief  at $\theta^\circ_s$. In summary, we arrive at the following conclusion.
 \begin{lemma} For the scenario of uniform beliefs to be attainable, agent $k$ should be connected either to all sending sub-networks or to none of them. 
 \end{lemma}
\noindent We provide in reference [arXiv] two numerical examples that illustrate this construction.

 \subsection{Example 1}
 Consider the network shown in Fig. \ref{network.label2}. It consists of $N=8$ agents, two sending sub-networks and one receiving sub-network, with the following combination matrix:
 
 \begin{small}
 	\begin{equation}
 	A=\left[
 	\begin{array}{ccccc|ccc}
 	0.2		&	0.2 	     	&0.8		& 0 		&	0 		& 0		&0		&\times	 \\
 	0 .5   	&      0.4   		&0.1		& 0 		& 	0		&\times	&0		 &0 \\
 	0.3 		& 	0.4		&0.1		&0		& 	0 		&\times		&\times	&0	\\
 	0          	& 	0  		&0		& 0.4 	& 	0.3 		&\times 	&0		&\times	\\
 	0          	& 	0  		&0		& 0.6 	& 	0.7 		& 0	 	&\times		&0	\\
 	\hline
 	0          	& 	0  		&0		& 0 		& 	0		& 0.2	 	&0.3		&0.2	 \\
 	0		&      0		&0		& 0		& 	0		& 0.1		&0.2		&0.3	 \\
 	0          	& 	0  		&0		& 0 		& 	0		& 0.1		&0.2		&0.1	 \\
 	\end{array}
 	\right]
 	\label{label.eq11Ex1}\end{equation}
 \end{small}
  \begin{figure}[h!]
  	\centering
  	\includegraphics[scale=0.35]{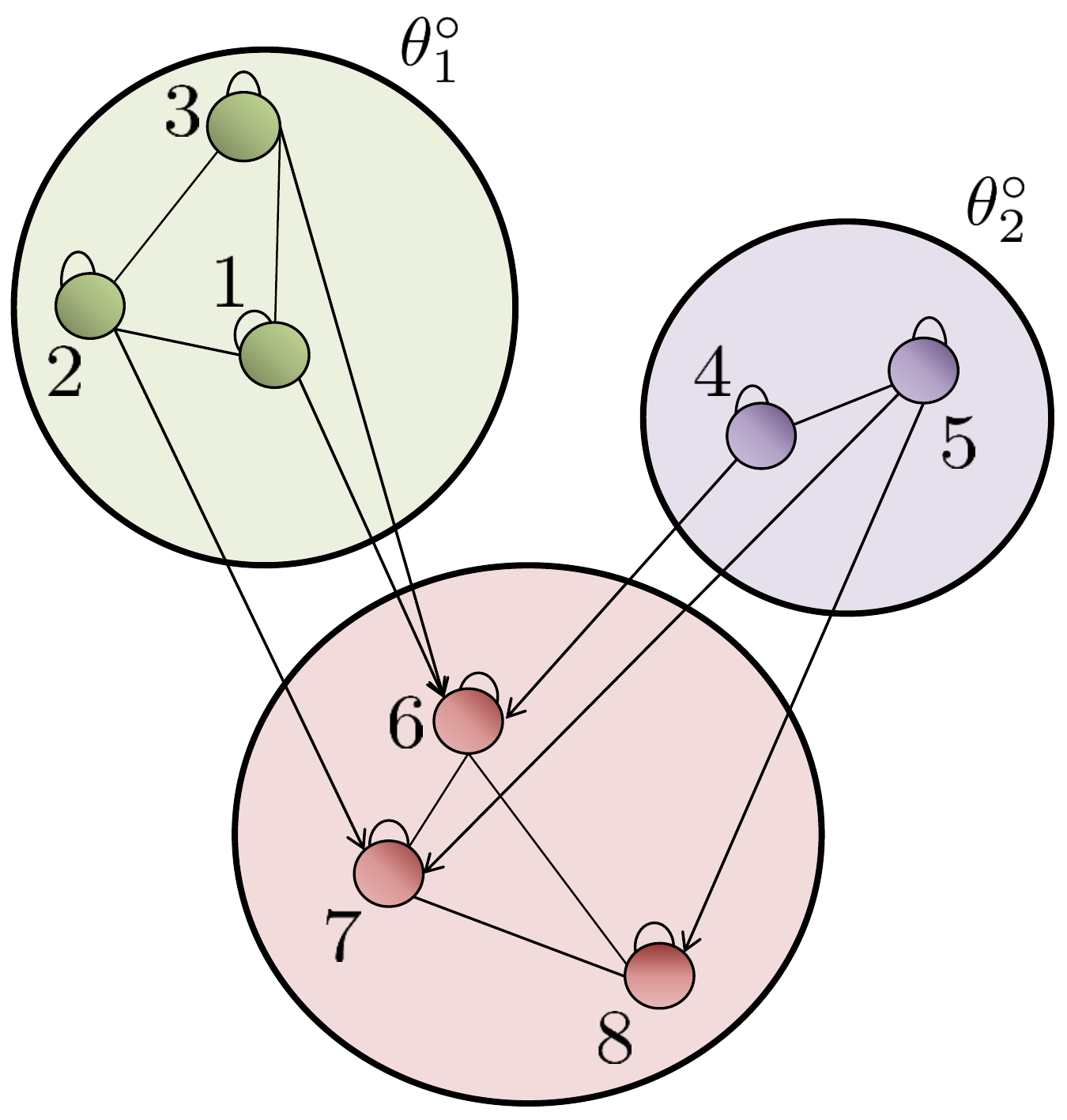} 
  	\caption{{\small A weakly connected network consisting of three sub-networks in a broadband influence scenario.}}
  	\label{network.label2}
  \end{figure}
  
 We assume that there are 3 possible states $\Theta=\{\theta_1^\circ,\theta_2^\circ,\theta_3^\circ\}$, where $\theta_1^\circ$ is the true event for the first sending sub-network, $\theta_2^\circ$ is the true event for the second sending sub-network, and $\theta_3^\circ$ is the true event for the receiving sub-network. 
 
  Let us first design $T_{SR}$ so that all receiving agents' beliefs converge to the same belief over $\{\theta_1^o,\theta_2^o\}$, say:
 \begin{align}
 q=\ba {c}
 0.2\\
 0.8
 \ea
 \end{align}
 We determine the columns of $T_{SR}$ one at a time. Starting with agent $6$, we focus on the first column of $T_{SR}$. The vector $v_6$ defined in (\ref{vk}) is given by (\ref{Uni}) for the case of uniform beliefs. Therefore,  
 \begin{align}
 v_6 &= 0.6q=\ba{c}
 0.12 \\
 0.48
 \ea
 \end{align}
 Thus, according to (\ref{tkFinal}),
 \begin{align}
 t_{SR,6}'&=\ba {c}
 \frac{v_6(1)}{2} \one_{2}\\
 v_6(2)
 \ea
 + 
 \ba {c}
 I_{2}-\frac{1}{2}\one_{2}\one_{2}\tran\\
1-\frac{1}{1} \one_{1}\one_{1}\tran \ea y
 \end{align}
 where $y$ is an arbitrary vector of length $3$. Note that $t'_{SR,6}$ represents respectively the coefficients of agents $2$, $3$ and $4$ that are linked to agent $6$. It follows that
 \begin{align}
 t_{SR,6}'
 &= \ba {c}
 0.06 \\
 0.06 \\
 0.48
 \ea
 + 
 \ba {c}
 \frac{1}{2}y(1)-\frac{1}{2}y(2) \\
 -\frac{1}{2}y(1)+\frac{1}{2}y(2) \\
 0
 \ea 
 \end{align}
  Let $\alpha_6\define \frac{1}{2}y(1)-\frac{1}{2}y(2)$ so that
 \begin{align}
 t_{SR,6}'=
 \ba {c}
 0.06+\alpha_6 \\
 0.06-\alpha_6 \\
 0.48
 \ea
 \end{align}
 In order to have positive entries for $t_{SR,6}'$, we can choose $|\alpha_6|\leq 0.06$. 
 
 Now for agent $7$, the vector $v_7$ is given by
 \begin{align}
 v_7 &= 0.3q=\ba{c}
 0.06 \\
 0.24
 \ea
 \end{align}
 
 so that
 \begin{align}
 t_{SR,7}'&=\ba {c}
 v_7(1)\\
 v_7(2)
 \ea =\ba {c}
 0.06\\
 0.24 
 \ea
 \end{align}
 The entries $t_{SR,7}'$ represent respectively the coefficients of agents $3$ and $5$ that are linked to agent $7$.

     Next for agent $8$, we have
     \begin{align}
     v_8 &=0.4q=\ba{c}
     0.08 \\
     0.32
     \ea
     \end{align}
     
     so that
     \begin{align}
     t_{SR,8}'&=\ba {c}
    { v_8(1)}\\
     {v_8(2)}
     \ea =\ba {c}
     0.08\\
     0.32
     \ea
     \end{align}
    
  Therefore, one possible solution is 
  \begin{align}
  T_{SR}=\ba{ccc}
  0 & 0 & 0.08 \\
  0.06 & 0 & 0 \\
  0.06 & 0.06 & 0 \\
  0.48 & 0 & 0.32 \\
  0    & 0.24 & 0
  \ea
  \end{align}
   To verify that the beliefs of the receiving agents converge in this case to the desired belief, we compute the matrix $W^{\sf T}$ from (\ref{defW1}):
   \begin{align}
   W\tran&=(I-T_{RR}\tran)^{-1}T_{SR}\tran \nn \\
   &=\ba{ccc c cc}
   0.0169 &  0.0839 & 0.0992 & & 0.7390 & 0.0610 \\
   0.0322 &  0.0394 & 0.1284 & & 0.4441 & 0.3559 \\
   0.1034 &  0.0318 & 0.0648 & & 0.6678 & 0.1322
   \ea
   \end{align}
   Then, according to (\ref{FinalDistribution}), we can compute the belief at $\theta_1^\circ$ for each receiving agent at steady state, by taking the first block in the agent's corresponding row and summing its elements:
   $$\lim_{i\to\infty}\bm \mu_{k,i}(\theta_1^\circ)=\left\{
   \begin{aligned}
   0.0169 +0.0839+0.0992=0.2,\ &\quad k=6\\
   0.0322+0.0394+0.1284=0.2,\ &\quad  k=7\\
   0.1034+0.0318+0.0648=0.2,\ &\quad  k=8
   \end{aligned}\right.$$
   Likewise, we can compute the belief at $\theta_2^\circ$ for each receiving agent at steady state, by taking the second block in the agent's corresponding row and summing its elements:
   $$\lim_{i\to\infty}\bm \mu_{k,i}(\theta_2^\circ)=
   \left\{
   \begin{aligned}
   0.7390+0.0610=0.8,\ &\quad k=6\\
   0.4441+0.3559=0.8,\ &\quad  k=7\\
   0.6678+0.1322=0.8,\ &\quad  k=8
   \end{aligned}\right.$$
   Let us now consider the case where we want to design $T_{SR}$ so that the desired limiting beliefs are not necessarily uniform but rather
   \begin{align}
    Q=\ba{c c c}
    0.8 & 0.7 & 0.75 \\
    0.2 & 0.3 & 0.25
    \ea
    \end{align}
    Note that now the beliefs are different from an agent to another, but they are still close. 
    Computing,
    \begin{align}
    v_k=q_k-Qt_{RR,k}
    \end{align}
    for each receiving agent $k$, we obtain:
   \begin{align}
   v_6 &= q_6-Qt_{RR,6}=\ba{c}
   0.495 \\
   0.105\ea \\
     v_7 &= q_7-Qt_{RR,7}=\ba{c}
     0.17 \\
     0.13
     \ea\\
 v_8 &= q_8-Qt_{RR,8} =\ba{c}
       0.305 \\
       0.195
       \ea
       \end{align}
       Therefore, one possible $T_{SR}$ is
       \begin{align}
       T_{SR}=\ba{ccc}
       0 & 0 & 0.305 \\
       0.495/2 & 0 & 0 \\
       0.495/2 & 0.17 & 0 \\
       0.105 & 0 & 0.195 \\
       0    & 0.13 & 0
       \ea
       \end{align} 
   Let us now consider the case where the desired limiting beliefs are more dispersed, such as
   \begin{align}
   Q=\ba{c c c}
   0.8 & 0.2 & 0.3 \\
   0.2 & 0.8 & 0.7
   \ea
   \end{align}
   In this case for agent $7$, we have
   \begin{align}
   v_7 &= q_7-Qt_{RR,7} =\ba{c}
   -0.14 \\
   0.44
   \ea
   \end{align}
   with a negative first entry. Therefore, the desired belief for agent $7$ cannot be attained. 
   
   \subsection{Example 2}
   Consider now the network shown in Fig. \ref{network.label} with the following combination matrix
    \begin{small}
     	\begin{equation}
     	A=\left[
     	\begin{array}{ccccc|ccc}
     	0.2		&	0.2 	     	&0.8		& 0 		&	0 		& 0		&0		&\times	 \\
     	0 .5   	&      0.4   		&0.1		& 0 		& 	0		&\times	&0		 &\times \\
     	0.3 		& 	0.4		&0.1		&0		& 	0 		&\times		&\times		&0	\\
     	0          	& 	0  		&0		& 0.4 	& 	0.3 		&\times 	&0		&0	\\
     	0          	& 	0  		&0		& 0.6 	& 	0.7 		& 0	 	&\times		&0	\\
     	\hline
     	0          	& 	0  		&0		& 0 		& 	0		& 0.2	 	&0.3		&0.1	 \\
     	0		&      0		&0		& 0		& 	0		& 0.1		&0.2		&0.6	 \\
     	0          	& 	0  		&0		& 0 		& 	0		& 0.1		&0.2		&0	 \\
     	\end{array}
     	\right]
     	\label{label.eq11Ex2}\end{equation}
     \end{small}
      \begin{figure}[h]
      	\centering
      	\includegraphics[scale=0.35]{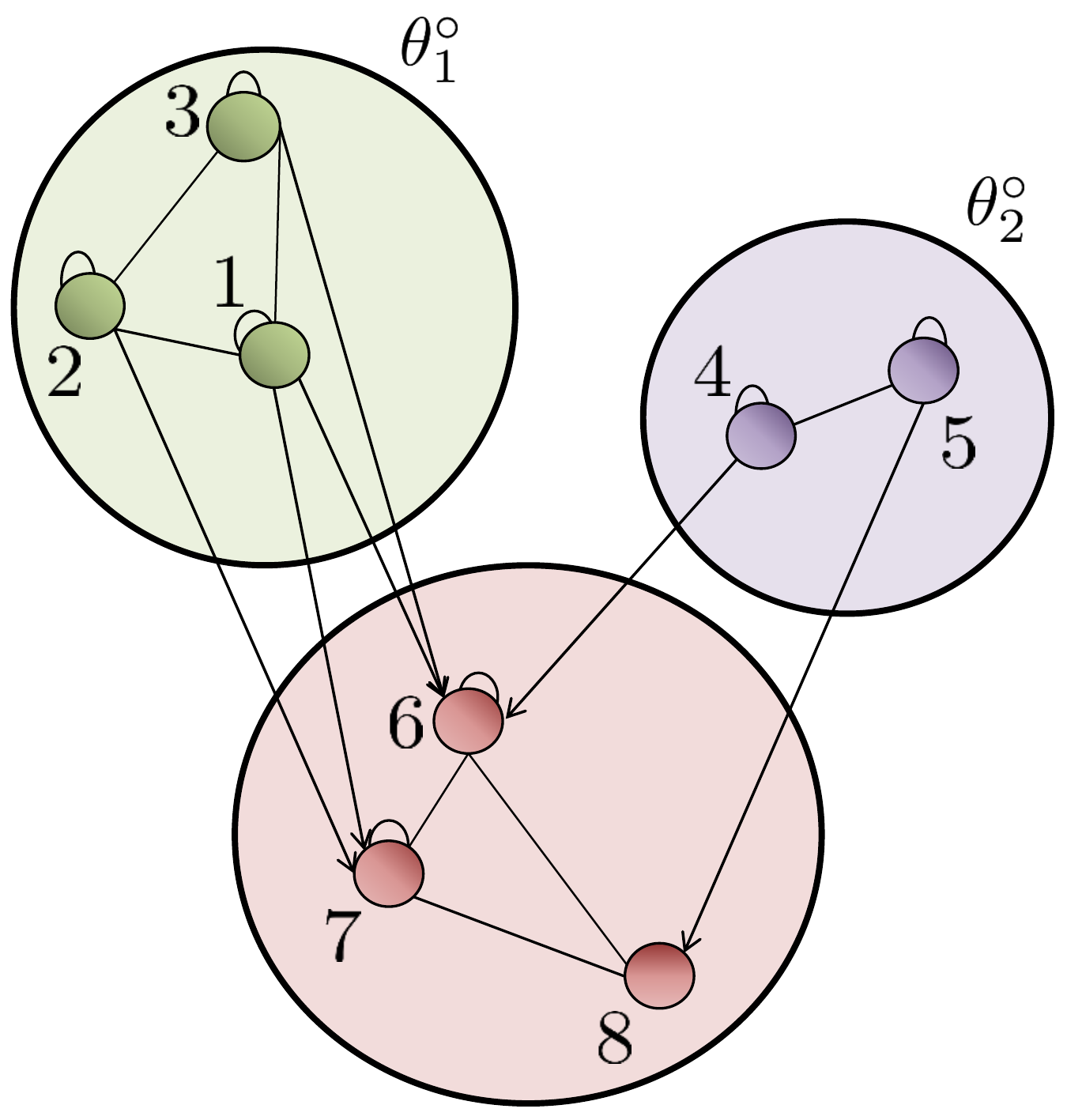} 
      	\caption{{\small A weakly connected network consisting of three sub-networks.}}
      	\label{network.label}
      \end{figure}
      
    \noindent Let us consider the case where we want to design $T_{SR}$ so that the desired limiting beliefs are as follows:
      \begin{align}
      Q=\ba{c c c}
      0.8 & 0.7 & 0.8 \\
      0.2 & 0.3 & 0.2
      \ea
      \end{align}
       Computing,
       \begin{align}
       v_k=q_k-Qt_{RR,k}
       \end{align}
       for each receiving agent $k$, we obtain:
       \begin{align}
       v_6 &= q_6-Qt_{RR,6}=\ba{c}
       0.49 \\
       0.11
       \ea  \\
       v_7 &= q_7-Qt_{RR,7} =\ba{c}
       0.16 \\
       0.14
       \ea  \\
       v_8 &= q_8-Qt_{RR,8} =\ba{c}
       0.3\\
       0
       \ea
       \end{align}
       Note that in this example, agent $8$ is not connected to the second sending sub-network, but the controlling scheme can still work because condition (\ref{noCon}) is satisfied. Therefore, one possible choice for $T_{SR}$ is the following:
        \begin{align}
        T_{SR}=\ba{ccc}
        0 & 0 & 0.3/2 \\
        0.49/2 & 0 & 0.3/2 \\
        0.49/2 & 0.16 & 0 \\
        0.11 & 0 & 0 \\
        0    & 0.14 & 0
        \ea
        \end{align}
        To verify that the beliefs of the agents converge in this case to the desired belief, we compute $W^{\sf T}$ from (\ref{defW1}) and use (\ref{FinalDistribution}) to determine the limiting beliefs at $\theta_1^o$ and $\theta_2^o$ at the receiving agents. This calculation gives
        $$\lim_{i\to\infty}\bm \mu_{k,i}(\theta_1^\circ)=\left\{
        \begin{aligned}
         0.0309 +0.3737 + 0.3954  =0.8,\ &\quad k=6\\
         0.0586  + 0.2200  + 0.4214=0.7,\ &\quad  k=7\\
         0.1883  +0.3193   + 0.2924=0.8,\ &\quad  k=8
        \end{aligned}\right.$$
     and
        $$\lim_{i\to\infty}\bm \mu_{k,i}(\theta_2^\circ)=
        \left\{
        \begin{aligned}
         0.1539  + 0.046=0.2,\ &\quad k=6\\
          0.0724   +0.2276=0.3,\ &\quad  k=7\\
        0.0588   + 0.1412=0.2,\ &\quad  k=8
        \end{aligned}\right.$$

  \section{Joint Design of $T_{RR}$ and $T_{SR}$}
In the previous sections, we analyzed the conditions that drive receiving agents to desired beliefs. The approach relies on determining the entries of the weighting matrix $T_{SR}$ from knowledge
of $Q$ (the desired beliefs) and $T_{RR}$ (the internal weighting structure within the receiving sub-networks). We saw how there is limitation to where the beliefs of receiving agents can converge. In particular, the internal combination of receiving sub-networks contribute to this limitation. We now examine the problem of designing $T_{SR}$ and $T_{RR}$ jointly, to see whether by having more freedom in choosing the coefficients of $T_{RR}$, we still encounter limitations on how to influence the receiving agents. We assume that we know the number of receiving sub-networks and the number of agents in each of these sub-networks. Using (\ref{condition}), we have
  \begin{align}
 \underbrace{ \ba{cc}
  E & Q
  \ea}_{\define B}
  \ba{c}
  T_{SR}\\
  T_{RR}
  \ea
  =Q \label{condition2}
  \end{align}
   Therefore, given $Q$ (the desired limiting beliefs of the receiving agents), the design problem becomes one of finding matrices $T_{SR}$ and $T_{RR}$ that satisfy (\ref{condition2}) subject to the following constraints: 
   \begin{align}
   \one\tran T_{SR}+&\one\tran T_{RR}=\one\tran \label{condd1}\\
   T_{SR,k}(j)&=0 \text{, if sending agent $j$ does not feed into $k$} \nn \\ T_{SR,k}(j)&\geq 0\text{, otherwise} \label{condd3}\\
   T_{RR,k}(j)&=0 \text{, if receiving agent $j$ does not feed into $k$ }\nn \\
   T_{RR,k}(j)&> 0 \text{, otherwise} \label{condd5}
   \end{align}
   In the last condition (\ref{condd5}), we are requiring $T_{RR,k}(j)$ to be strictly positive if receiving agent $j$ feeds into $k$. This is in order to avoid solutions where the receiving sub-networks become  unconnected. For instance, consider the example shown in Fig. \ref{JointEx}. This figure shows a case where agent $k$ is connected to all sending sub-networks, and it depicts only the incoming links into agent $k$. Let us assume that the desired limiting belief for agent $k$ is
   \begin{align}
   \ba{c}
   q_k(\theta^\circ_1)\\
   q_k(\theta^\circ_2)
   \ea=
   \ba{c}
   0.1\\
   0.9
   \ea
   \end{align}
   Then a possible solution to (\ref{condition2}) is to assign zero as weights for the data originating from its receiving neighbors, $0.1$ for the data received from sending agent $1$, and $0.9$ for the data received from sending agent $2$. Then, for this example,
   \begin{align}
   E&=\ba{cc}
   1&0\\
   0&1
   \ea,\;Q=
   \ba{ccc}
   0.1&q_4(\theta^\circ_1)&q_5(\theta^\circ_1)\\
   0.9&q_4(\theta^\circ_2)&q_5(\theta^\circ_2)
   \ea,\nn\\
   T_{SR}&=\ba{c}0.1\\0.9\ea,\;T_{RR}=\ba{c}0\\0\\0\ea
   \end{align}
   so that (\ref{condition2}) is satisfied. However, this solution affects the connectedness of the receiving sub-network of agent $k$, because there will be no path that leads to this agent.
    \begin{figure}[h!]
       \centering
       \includegraphics[scale=0.35]{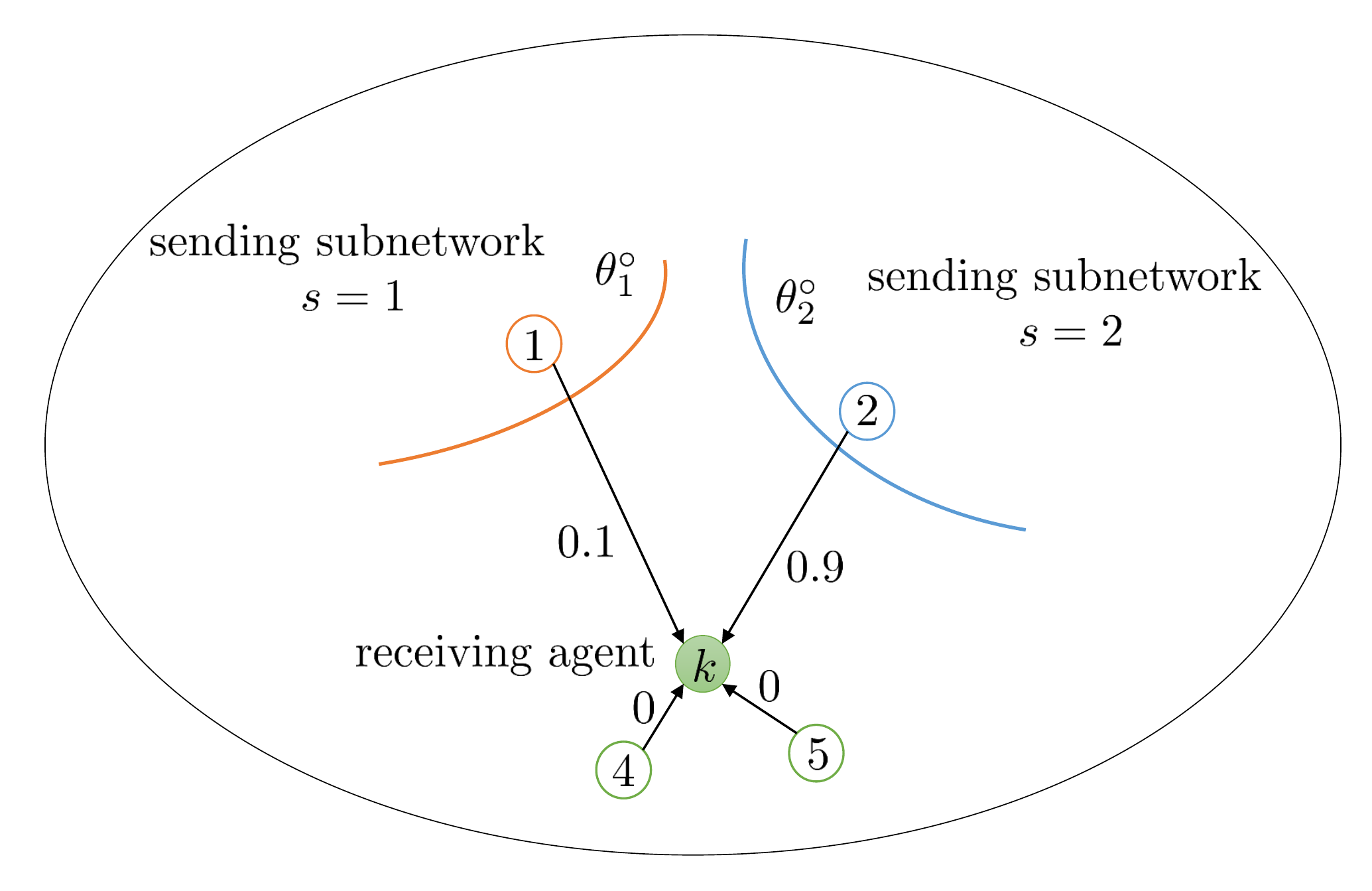} 
        	\caption{\small An example where the receiving network of agent $k$ ends up being disconnected.}
        	\label{JointEx}
        \end{figure}
   
   To find $T_{SR}$ and $T_{RR}$ satisfying (\ref{condition2})-(\ref{condd5}), we can solve separately for each of their columns. If it is possible to find a solution for each column, then $Q$ is attainable. We explore next the possibility of finding solutions for each column. Similarly to the previous section, $t_{SR,k}$ and $t_{RR,k}$ respectively represent the columns of $T_{SR}$ and $T_{RR}$ that correspond to receiving agent $k$, and $t_{SR,k}(j)$ and $t_{
   	RR,k}(j)$ respectively represent the $j-$th entries of this $t_{SR,k}$ and $t_{RR,k}$. Also $q_k$ denote the column of $Q$ that corresponds to receiving agent $k$. Then, relations (\ref{condition2}) and (\ref{condd3})--(\ref{condd5}) imply that the columns $t_{SR,k}$ and $t_{RR,k}$ must satisfy:
    \begin{align}
    \underbrace{ \ba{cc}
    	E & Q
    	\ea}_{= B}
    \ba{c}
    t_{SR,k}\\
    t_{RR,k}
    \ea
    =q_k \label{condition2vec}
    \end{align}
    subject to
    
    \begin{align}
    	\one\tran t_{SR,k}+&\one\tran t_{RR,k}=1 \label{condd1vec}\\
    	t_{SR,k}(j)&=0 \text{, if $j$ does not feed $k$} \nn \\ t_{SR,k}(j)&\geq 0\text{, otherwise} \label{condd3vec}\\
    	t_{RR,k}(j)&=0 \text{, if $j$ does not feed $k$ }\nn \\
    	t_{RR,k}(j)&>0 \text{, otherwise} \label{condd5vec}
    \end{align}
   
   Since the connections within the sending and receiving networks are known, but not the combination weights $T_{SR}$ and $T_{RR}$ whose values we are seeking, we can then set to zero the entries of $t_{SR,k}$ and $t_{RR,k}$ that correspond to unlinked agents. We remove these zero entries and relabel the vectors as $t_{SR,k}'$ and $t_{RR,k}'$. We also remove the corresponding columns of $E$ and $Q$, and label the modified $E$ and $Q$ by $E_k$ and $Q_k$. We are therefore reduced to determining $t_{SR,k}'$ and $t_{RR,k}'$ by solving a system of equations of the form:
    \begin{align}
    \underbrace{ \ba{cc}
    	E_k & Q_k
    	\ea}_{\define B_k}
    \ba{c}
    t_{SR,k}'\\
    t_{RR,k}'
    \ea
    =q_k \label{condition2vecR}
    \end{align}
    subject to
    \begin{align}
    \one\tran t_{SR,k}'+&\one\tran t_{RR,k}'=1 \label{finalCond1}\\
    t'_{SR,k}&\succcurlyeq \boldsymbol{0} \label{finalCond2}\\
    t'_{RR,k}&\succ\boldsymbol{0} \label{finalCond3}
    \end{align} 
    Formulation (\ref{condition2vecR})-(\ref{finalCond3}) has the following interpretation. After some sufficient time $i\geq I$, we know that the beliefs of all agents will approach some limiting beliefs, and based on the results of the previous work \cite{salamijournal}, the belief update (\ref{eqn:diffusion}) approaches for $i\geq I$,
    \be
    \left\{
    \begin{aligned}
    	\bm\psi_{k,i+1}(\theta) & =\bmu_{k,i}(\theta) \\ 
    	\bmu_{k,i+1}(\theta) & =\sum_{\ell\in\mathcal{N}_k}a_{\ell k}\,\bm\psi_{\ell,i+1}(\theta)=\sum_{\ell\in\mathcal{N}_k}a_{\ell k}\,\bm\mu_{\ell,i}(\theta) 
    \end{aligned}\right.
    \ee
    This means that:
    \begin{align}
    	\lim_{i\to\infty}\bmu_{k,i}(\theta) & =\sum_{\ell\in\mathcal{N}_k}a_{\ell k}\,\left(\lim_{i\to\infty}\bm\mu_{\ell,i}(\theta) \right)
    \end{align}
    In other words, if we want the beliefs of the receiving agents to converge to some belief vector $q$, then we need to make sure that these desired beliefs satisfy the relationship:
    \begin{align}
    	q_{k}(\theta) & =\sum_{\ell\in\mathcal{N}_k}a_{\ell k} q_{\ell}(\theta) \label{roula}
    \end{align}
    for any $\theta \in \{\theta^\circ_1,\cdots,\theta^\circ_S\}$ and for all receiving agents $k$. In other words, given the set of desirable beliefs, we would like to know if it is possible to express the desired limiting belief for each receiving agent $k$ as a convex combination of the limiting beliefs of its receiving neighbors and the limiting beliefs of the sending agents to which agent $k$ is connected. If this is possible for each agent $k$, then $Q$ is attainable, i.e., all receiving agents can reach their desired limiting beliefs. This is precisely what the formulation (\ref{condition2vecR})--(\ref{finalCond3}) is attempting to enforce, by finding suitable coefficients such that (\ref{roula}) is satisfied.
     Finding $t'_{SR,k}$ and $t'_{RR,k}$ that satisfy (\ref{condition2vecR}) and constraints (\ref{finalCond1})-(\ref{finalCond3}) might not be always possible. Since each agent $k$ can be connected to all sending sub-networks, or to some of them or to none of them, the matrix $E_k$ that appears in (\ref{condition2vecR}) will have a different form for each of these cases, which will affect the possibility of finding a solution. Before analyzing how the three possible cases affect the possibility of finding a solution, we summarize first the results:
     \begin{enumerate}
     	\item Agent $k$ is connected to all sending sub-networks: the problem reduces to finding $t'_{RR,k}$ that satisfies (\ref{condi1}) and (\ref{condi2}, which always has a solution;
     	\item Agent $k$ is connected to some sending sub-networks: the problem  reduces to finding $t'_{RR,k}$ that satisfies conditions (\ref{condi2-1})-(\ref{condi2-3}), which may not always have a solution;
     	\item Agent $k$ is not connected to any sending sub-network:  the problem reduces to finding $t'_{RR,k}$ that satisfies conditions (\ref{case3})-(\ref{case3cond}), which may not always have a solution.
     \end{enumerate} 
 
      Note that relations (\ref{condi1}) and (\ref{condi2-1}) are what condition (\ref{wtfCon}) required when we wanted to design $T_{SR}$, for the case where agent $k$ is connected to sending sub-network $s$, when $T_{RR}$ was given. 
      Similarly, relations (\ref{condi2-2}) and (\ref{case3}) are what condition (\ref{noCon}) required when we wanted to design $t_{SR,k}$, for the case where agent $k$ is not connected to sending sub-network $s$, when $T_{RR}$ was given. In the earlier section, we had to make sure that the given $T_{RR}$ satisfies (\ref{noCon}) and (\ref{wtfCon}) for $Q$ to be attainable. Here, we are designing for $T_{RR}$ as well, and we need to make sure that the entries we choose satisfy these conditions. We now analyze each case in detail.
      
    \subsection*{Case 1: Agent $k$ is connected to all sending sub-networks}
    We discuss first the case where agent $k$ is connected to at least one agent from each sending sub-network. In this case, $E_k$ will have the following form:
    \begin{align}
   E_k=	\ba {c c c c}
   \one_{N_1^k}\tran&\boldsymbol{0}&\hdots&\boldsymbol{0}\\
   \boldsymbol{0}&\one_{N_2^k}\tran&\hdots&\boldsymbol{0}\\
   \vdots& \vdots&\ddots &\vdots \\
   \boldsymbol{0}&\boldsymbol{0}&\hdots&\one_{N_{S}^k}\tran
   \ea
    \end{align}
  and relation (\ref{condition2vecR}) is then:
  \begin{align}
  \ba {ccccccc}
  \one_{N_1^k}\tran&\boldsymbol{0}&\hdots&\boldsymbol{0}&q_{k(1)}(\theta^\circ_1)&\hdots&q_{k(N_{gR}^k)}(\theta^\circ_1)\\
  \boldsymbol{0}&\one_{N_2^k}\tran&\hdots&\boldsymbol{0}&q_{k(1)}(\theta^\circ_2)&\hdots&q_{k(N_{gR}^k)}(\theta^\circ_2)\\
  \vdots& \vdots&\ddots &\vdots&\vdots&\ddots&\hdots \\
  \boldsymbol{0}&\boldsymbol{0}&\hdots&\one_{N_{S}^k}\tran&q_{k(1)}(\theta^\circ_s)&\hdots&q_{k(N_{gR}^k)}(\theta^\circ_s)
  \ea& \nn\\
  \ba{c}
  t_{SR,k}'\\t_{RR,k}'
  \ea
  =q_k \label{relCase1}
  \end{align}
     where $q_{k(j)}(\theta^\circ_s)$ represents the desired limiting belief at $\theta^\circ_s$ for the $j^{th}$ receiving neighbor of agent $k$, and $N^k_{gR}$ is the total number of receiving agents that are neighbors of agent $k$. The problem here is to find $t_{SR,k}'$ and $t_{RR,k}'$ that satisfy (\ref{relCase1}) subject to the constraints (\ref{finalCond1})-(\ref{finalCond2}). It is useful to note that if we can find $t_{SR,k}'$ and $t_{RR,k}'$ that satisfy (\ref{relCase1}), then condition (\ref{finalCond1}) will be automatically satisfied. To see this, we first sum the elements of the vector on the left-hand side of (\ref{relCase1}) and observe that
   \begin{align}
   \one\tran_{S}B_k \ba{c}t_{SR,k}'\\t_{RR,k}'\ea=\one\tran\ba{c}t_{SR,k}'\\t_{RR,k}'\ea \label{rell1Case1}
   \end{align}
  The matrix $B$ was introduced in (\ref{condition2vec}). This is because $\one_S\tran B_k=\one\tran$ since the entries on each column of $B_k$ add up to one. We then sum the elements of the vector on the right-hand side of (\ref{relCase1}) to get
   \begin{align}
   \one_{S}\tran q_k=1 \label {rell2Case1}
   \end{align}
  Thus, equating (\ref{rell1Case1}) and (\ref{rell2Case1}), we obtain (\ref{finalCond1}). The problem we are attempting to solve is then equivalent to finding $t_{SR,k}'$ and $t_{RR,k}'$ that satisfy (\ref{relCase1}) subject to
     \begin{align}
     t'_{SR,k}&\succcurlyeq \boldsymbol{0} \label{finalCond2Eq}\\
     t'_{RR,k}&\succ\boldsymbol{0} \label{finalCond3Eq}
     \end{align} 
  Now, note that (\ref{relCase1}) consists of $S$ equations and note that the number of variables (i.e., the total number of entries of $t'_{SR,k}$ and $t'_{RR,k}$) is greater than the number of equations. Each equation relates the entries of $t'_{SR,k}$ that correspond to agents of one of the sending sub-networks to all entries of $t'_{RR,k}$. In particular, the equation that corresponds to sending sub-network $s$ has the following form:
   \begin{align}
     \sum_{\ell\in\mathcal{I}_s} t_{SR,k}'(\ell)=q_k(\theta^\circ_s)-\sum_{j=1}^{N_{gR}^k}q_{k(j)}(\theta^\circ_s)t'_{RR,k}(j) \label{case1}
   \end{align}
 Equation (\ref{case1}) shows how the entries of $t'_{SR,k}$ that correspond to agents of sending sub-network $s$, are related to the entries of $t'_{RR,k}$  through the values of the desired beliefs at $\theta^\circ_s$. Therefore, the set of all possible solutions to (\ref{relCase1}) consist of vectors whose entries satisfy (\ref{case1}) for each $s$. In other words, by arbitrarily fixing the entries of $t'_{RR,k}$, we compute the entries of $t_{SR,k}'$ using (\ref{case1}) for each $s$ to obtain a solution to (\ref{relCase1}). This is because (\ref{relCase1}) is made of $S$ equations that only indicate how the entries of $t'_{SR,k}$ that correspond to each sending sub-network $s$ are related to $t'_{RR,k}$ without having any additional equation for the entries of $t'_{RR,k}$. Note that it does not matter how the individual entries of $t'_{SR,k}$ that correspond to sub-network $s$ are chosen as long as their sum satisfies (\ref{case1}). However, in the problem we are trying to solve, we are not interested in the entire set of solutions to (\ref{relCase1}). This is because we have two additional constraints (\ref{finalCond2Eq}) and (\ref{finalCond3Eq}). Therefore, in our problem we cannot arbitrarily fix  the entries of $t'_{RR,k}$ to any values as we need to also satisfy (\ref{finalCond2Eq}) and (\ref{finalCond3Eq}). Constraint (\ref{finalCond2Eq}) implies that (\ref{case1}) should be non-negative for each sending sub-network $s$, i.e.,
     \begin{align}
q_k(\theta^\circ_s)\geq\sum_{j=1}^{N_{gR}^k}q_{k(j)}(\theta^\circ_s)t'_{RR,k}(j) \label{case12}
     \end{align}
   Therefore, the problem reduces to finding $t'_{RR,k}$ that satisfies:
\begin{subequations}
	\begin{empheq}[box=\widefbox]{align}
	 q_k(\theta^\circ_s)&\geq\sum_{j=1}^{N_{gR}^k}q_{k(j)}(\theta^\circ_s)t'_{RR,k}(j),\;\;\; \forall s \label{condi1} \\
		  t'_{RR,k}&\succ 0   \label{condi2}
\end{empheq}
\end{subequations}
   If it possible to find $t'_{RR,k}$ that satisfies (\ref{condi1}) and (\ref{condi2}), then $t'_{SR,k}$ can de determined using (\ref{case1}) and therefore a solution for agent $k$ is found. Finding $t'_{RR,k}$ that satisfies (\ref{condi1}) and (\ref{condi2}) is always possible. By appropriately attenuating the entries of $t'_{RR,k}$, we can have the right-hand side of (\ref{condi1}) smaller than $q_k(\theta^\circ_s)$. For instance, one solution is to assign the same value  $\epsilon_k>0$ to all entries of $t_{RR,k}'$. Then from (\ref{condi1}), we have for each $s$:
     \begin{align}
     q_k(\theta^\circ_s)\geq\epsilon_k\sum_{j=1}^{N_{gR}^k}q_{k(j)}(\theta^\circ_s) \label{case13}
     \end{align}
     which means that $\epsilon_k$ should be chosen so that:
    \begin{align}
      0<\epsilon_k\leq \min_s \left\{ \frac{q_k(\theta^\circ_s)}{\sum_{j=1}^{N_{gR}^k}q_{k(j)}(\theta^\circ_s)}\right\} \label{epsi1}
    \end{align}
  
  We mentioned that, after finding $t'_{RR,k}$ that satisfies (\ref{condi1}) and (\ref{condi2}), $t'_{SR,k}$ can be determined using (\ref{case1}). We can alternatively express the solutions of $t'_{SR,k}$ using the same approach of the previous section. This is because after choosing the entries of $t'_{RR,k}$, the problem is now similar to the previous problem of finding $t_{SR,k}$ while $t_{RR,k}$ is given. Therefore, the solutions for $t'_{SR,k}$ can be also given by (\ref{tkFinal}). Note that (\ref{tkFinal}) is expressed in terms of $v'_{k}$ to take into account that agent $k$ may not be connected to some sending sub-networks, in the earlier section. Since in this case we are focusing on agent $k$ connected to all sending sub-networks, the solution for $t'_{SR,k}$ is given by (\ref{tkFinal} where $v_{k}$ is used instead of  $v'_{k}$.

In summary, when agent $k$ is connected to all sending sub-networks, the problem can have an infinite number of solutions. We first find $t'_{RR,k}$ that satisfies (\ref{condi1}) and (\ref{condi2}). Then, the entries of $t_{SR,k}'$ are nonnegative values chosen to satisfy (\ref{case1}). In other words, when a receiving agent $k$ is under the direct influence of all sending sub-networks, it is relatively straightforward to affect its beliefs, especially since the influence from its receiving neighbors can be attenuated as much as needed through the choice $\epsilon_k$.

    \subsection*{Case 2: Agent $k$ is connected to some sending sub-networks}
    We now consider the case where agent $k$ is influenced by only a subset of the sending networks. Without loss of generality, we assume it is connected to the first $s'$ sending sub-networks. In this case, $E_k$ will have the following form:
    \begin{align}
     E_k= \ba {c c c c}
     \one_{N_1^k}\tran&\boldsymbol{0}&\hdots&\boldsymbol{0}\\
     \boldsymbol{0}&\one_{N_2^k}\tran&\hdots&\boldsymbol{0}\\
     \vdots& \vdots&\ddots &\vdots \\
     \boldsymbol{0}&\boldsymbol{0}&\hdots&\one_{N_{s'}^k}\tran\\
     \boldsymbol{0}&\boldsymbol{0}&\hdots&\boldsymbol{0}\\
     \vdots& \vdots&\vdots &\vdots \\
     \boldsymbol{0}&\boldsymbol{0}&\hdots&\boldsymbol{0}
     \ea
    \end{align}
and relation (\ref{condition2vecR}) becomes:
     \begin{align}
     \ba {c c c c c c c}
     \one_{N_1^k}\tran&\boldsymbol{0}&\hdots&\boldsymbol{0}&q_{k(1)}(\theta^\circ_1)&\hdots&q_{k(N_{gR}^k)}(\theta^\circ_1)\\
     \boldsymbol{0}&\one_{N_2^k}\tran&\hdots&\boldsymbol{0}&q_{k(1)}(\theta^\circ_2)&\hdots&q_{k(N_{gR}^k)}(\theta^\circ_2)\\
     \vdots& \vdots&\ddots &\vdots&\vdots&\ddots&\vdots \\
     \boldsymbol{0}&\boldsymbol{0}&\hdots&\one_{N_{s'}^k}\tran&q_{k(1)}(\theta^\circ_{s'})&\hdots&q_{k(N_{gR}^k)}(\theta^\circ_{s'})\\
     \boldsymbol{0}&\boldsymbol{0}&\hdots&\boldsymbol{0}&q_{k(1)}(\theta^\circ_{{s'}+1})&\hdots&q_{k(N_{gR}^k)}(\theta^\circ_{{s'}+1})\\
     \vdots& \vdots&\vdots &\vdots&\vdots&\vdots&\vdots \\
     \boldsymbol{0}&\boldsymbol{0}&\hdots&\boldsymbol{0}&q_{k(1)}(\theta^\circ_S)&\hdots&q_{k(N_{gR}^k)}(\theta^\circ_S)\\
     \ea& \nn \\
     \ba{c}
     t_{SR,k}'\\ t_{RR,k}'\ea
     =q_k \label{relCase2}
     \end{align}
     The problem now is to find $t_{SR,k}'$ and $t_{RR,k}'$ that satisfy (\ref{relCase2}) subject to constraints (\ref{finalCond1})-(\ref{finalCond2}). 
     As before, if we can find $t_{SR,k}'$ and $t_{RR,k}'$ that satisfy (\ref{relCase2}), then condition (\ref{finalCond1}) will be automatically satisfied. Note now that (\ref{relCase2}) consists of $s'$ equations that relate the entries of $t'_{SR,k}$ to the entries of $t'_{RR,k}$, and $S-s'$ equations that involve the entries of $t'_{RR,k}$. Therefore, any vector that satisfies (\ref{relCase2}) will have the following property:
     \begin{align}
     \sum_{\ell=1 }^{N_1^k} t_{SR,k}'(\ell)=q_k(\theta^\circ_s)-\sum_{j=1}^{N_{gR}^k}q_{k(j)}(\theta^\circ_s)t'_{RR,k}(j) \label{case2}
     \end{align} 
    but only for $s\leq s'$. In other words, the entries of $t'_{SR,k}$ that correspond to sub-network $s\leq s'$ are expressed in terms of $t'_{RR,k}$ through (\ref{case2}). In addition, and differently from case 1, any solution to (\ref{relCase2}) should also satisfy:
      \begin{align}
 \sum_{j=1}^{N_{gR}^k}q_{k(j)}(\theta^\circ_s)t'_{RR,k}(j)=   q_k(\theta^\circ_s) \label{case21}
      \end{align}
      for any $s>s'$. Likewise, constraint (\ref{finalCond2Eq}) implies that (\ref{case2}) should be non-negative for each sending sub-network $s$ where $s\leq s'$, i.e.,
      \begin{align}
      q_k(\theta^\circ_s)\geq\sum_{j=1}^{N_{gR}^k}q_{k(j)}(\theta^\circ_s)t'_{RR,k}(j) \label{case22}
      \end{align}
      for any $s\leq s'$. Therefore, the problem reduces to finding $t'_{RR,k}$ that satisfies:
  \begin{subequations}
  	\begin{empheq}[box=\widefbox]{align}
  	    q_k(\theta^\circ_s)&\geq\sum_{j=1}^{N_{gR}^k}q_{k(j)}(\theta^\circ_s)t'_{RR,k}(j),\;\;\; s\leq s' \label{condi2-1} \\
  	q_k(\theta^\circ_s)&=\sum_{j=1}^{N_{gR}^k}q_{k(j)}(\theta^\circ_s)t'_{RR,k}(j),\;\;\; s> s' \label{condi2-2} \\
  	t'_{RR,k}&\succ 0   \label{condi2-3} 
  	\end{empheq}
  \end{subequations}
     
       If it possible to find $t'_{RR,k}$ that satisfies (\ref{condi2-1})-(\ref{condi2-3}), then $t'_{SR,k}$ can de determined using (\ref{case2}) or alternatively using (\ref{tkFinal}). However, in contrast to the case studied in the previous case, finding $t'_{RR,k}$ that satisfies conditions (\ref{condi2-1})-(\ref{condi2-3}) may not be always possible. For instance, consider agent $k$ shown in Fig. \ref{Ex}, which is connected to only the first sending sub-network but not to the other two sending sub-networks. 
      \begin{figure}[h!]
      	\centering
      	\includegraphics[scale=0.35]{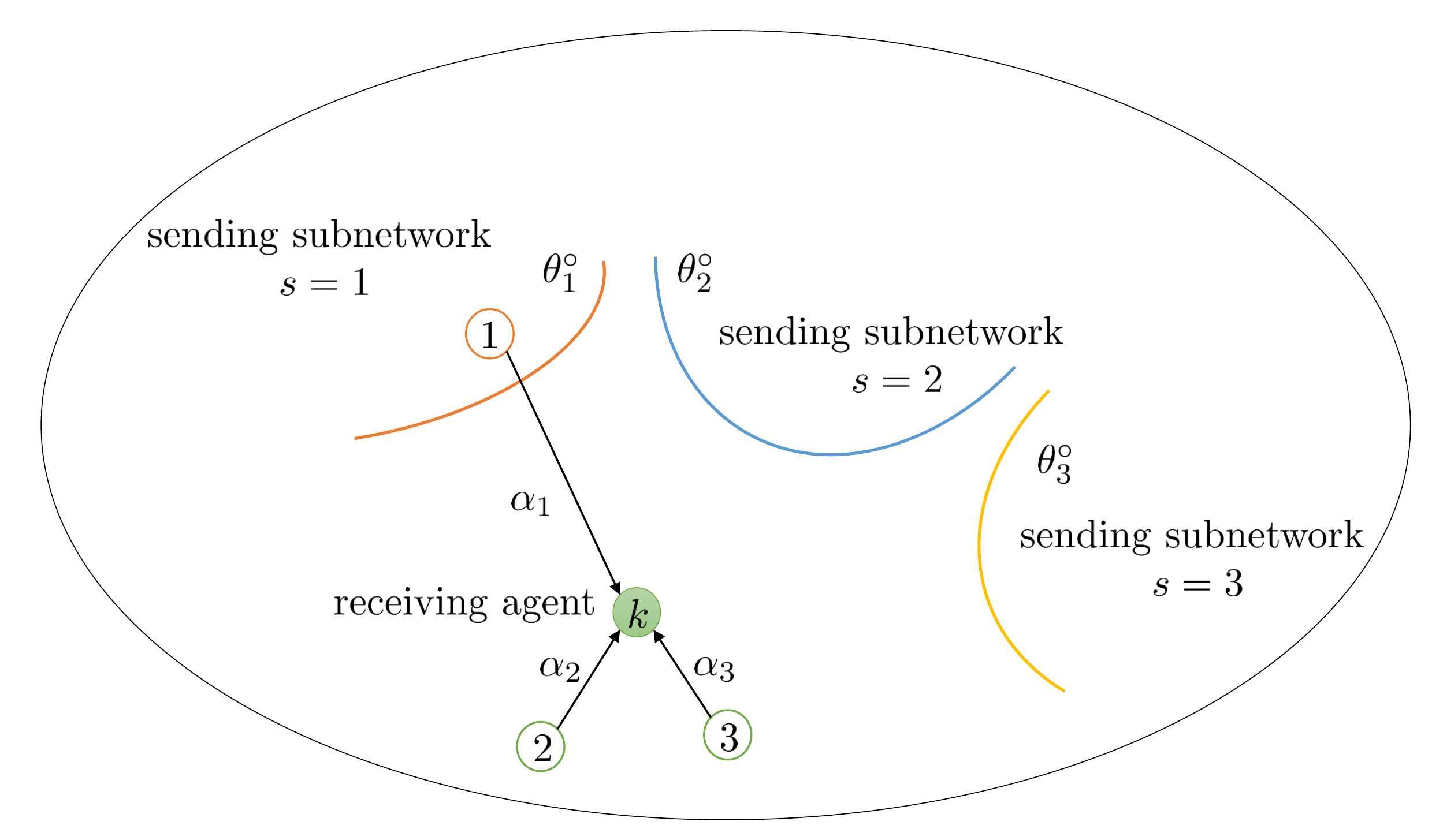} 
      	\caption{{\small An example where receiving agent $k$ is only connected to one sending sub-network.}}
      	\label{Ex}
      \end{figure}
       Let us consider its desired limiting belief as
      \begin{align}
      \ba{c}
      q_k(\theta^\circ_1)\\
      q_k(\theta^\circ_2)\\
      q_k(\theta^\circ_3)
      \ea= \ba{c}
      0.1\\
      0.45\\
      0.45
      \ea
      \end{align}
      while the desired limiting beliefs for its neighbors are:
        \begin{align}
        \ba{c}
        q_2(\theta^\circ_1)\\
        q_2(\theta^\circ_2)\\
        q_2(\theta^\circ_3)
        \ea= \ba{c}
        0.2\\
        0.5\\
        0.3
        \ea, \;\;
           \ba{c}
           q_3(\theta^\circ_1)\\
           q_3(\theta^\circ_2)\\
           q_3(\theta^\circ_3)
           \ea= \ba{c}
           0.1\\
           0.4\\
           0.5
           \ea
        \end{align}
      Then, from (\ref{condi2-1}), we should have:
      \begin{align}
       q_k(\theta^\circ_1)&\geq \alpha_2 q_2(\theta^\circ_1)+\alpha_3 q_3(\theta^\circ_1) \implies
       0.1\geq 0.2\alpha_2 +0.1\alpha_3 \label{check1} 
      \end{align}
      and from (\ref{condi2-2}),
       \begin{align}
       q_k(\theta^\circ_2)&= \alpha_2 q_2(\theta^\circ_2)+\alpha_3 q_3(\theta^\circ_2)\implies 0.45=0.5\alpha_2+0.4\alpha_3 \label{check2}\\
       q_k(\theta^\circ_3)&= \alpha_2 q_2(\theta^\circ_3)+\alpha_3 q_3(\theta^\circ_3)\implies 0.45=0.3\alpha_2+0.5\alpha_3 \label{check3}
       \end{align}
       Solving (\ref{check2}) and (\ref{check3}) gives the following solution: $\alpha_2= 0.3462$ and $\alpha_3=0.6923$. However,
       $0.2\alpha_2+0.1\alpha_3= 0.1385$, which violates (\ref{check1}). Still, we can have cases where all conditions (\ref{condi2-1})-(\ref{condi2-3}) can be met (we are going to provide one example in a later section), then in these cases, we choose $t_{SR,k}'$ according to (\ref{case2}). 

We observe from this case that the fewer the sending networks that influence agent $k$, the harder it is to affect its limiting belief. This emphasizes again the idea that the structure of the receiving sub-networks helps in limiting external manipulation.
       \subsection*{Case 3: Agent $k$ is not connected to any sending sub-networks}
       When agent $k$ is not connected to any sending sub-network, relation (\ref{condition2vecR}) reduces to:
       \begin{align}
         Q_k t_{RR,k}'=q_k 
       \end{align}
       The problem is then to find $t_{RR,k}'$ that satisfies:
       \begin{subequations}
       	\begin{empheq}[box=\widefbox]{align}
       	\	 Q_k t_{RR,k}'&=q_k \label{case3}\\
       	\one\tran t_{RR,k}'&=1\\
       	t_{RR,k}'&\succ 0\label{case3cond}
       	\end{empheq}
       \end{subequations}
       This problem might not have an exact solution. For instance, we discuss two examples in Appendix \ref{App.A} of [arXiv paper], where in the second example, we have an agent that is not connected to any sending sub-network and its desired belief cannot be expressed as a convex combination of the desired beliefs of its neighbors.  \\
       
%

       \noindent {\bf Comment and analysis}
       
       \noindent  Since the problem of finding $T_{SR}$ and $T_{RR}$ satisfying (\ref{condition2})-(\ref{condd5}) is separable, we studied the possibility of finding a solution for each column of $T_{SR}$ and $T_{RR}$. We analyzed the problem for 3 cases and discovered that for the first case (when agent $k$ is connected to at least one agent from each sending sub-network), problem (\ref{condition2vecR})--(\ref{finalCond3}) always has a solution. That is, if an agent $k$ is connected to all sending sub-networks and {\em given knowledge of the limiting beliefs of its neighbors}, we can always find the weight combination for agent $k$ such that (\ref{roula}) is satisfied. For the second case (when agent $k$ is connected to some sending sub-networks) and the third case (when agent $k$ is not connected to any sending sub-network), we found out that problem (\ref{condition2vecR})--(\ref{finalCond3}) might not always have a solution, i.e., it is not always possible to satisfy (\ref{roula}). These scenarios reinforce again the idea that the internal structure of receiving agents can resist some of the external influence.
       
       However, for $Q$ to be achievable (i.e., for the beliefs of all receiving agents converge to the desired beliefs), a solution must exist for each agent $k$. If the desired limiting belief of any receiving agent cannot be written as a convex combination of the limiting beliefs of its neighbors (i.e., a solution cannot be found for problem (\ref{condition2vecR})--(\ref{finalCond3})), the whole scenario is not achievable. Even if it is possible for agent $k$ to find its appropriate weights $t'_{SR,k}$ and $t'_{RR,k}$, finding this solution is based on the knowledge of the desired limiting beliefs of its neighbors. However, if one of the receiving neighbors cannot reach its desired belief, agent $k$ will not be able anymore to reach its desired belief. Therefore, for $Q$ to be attainable, a solution for problem (\ref{condition2vecR})--(\ref{finalCond3}) must exist for each receiving agent $k$. If $Q$ is not attainable, then the desired scenario should be modified to an attainable scenario, by taking into consideration the limitation provided by the internal connection of the receiving sub-networks. Or an approximate least-squares solution for the weights can be found. That is, we can instead seek to solve
             \begin{align}
             \min_{t_{SR,k}',t_{RR,k}'}\left\|B_k \ba{c}t_{SR,k}'\\t_{RR,k}'\ea-q_k \right\|^2 \label{prob}
             \end{align}
             subject to
             \begin{align}
              \one\tran t_{SR,k}'+\one\tran t_{RR,k}'=1 \label{cont1}\\
                  t_{SR,k}'\succeq \boldsymbol{0}\label{cont2}\\
                  t_{RR,k}'\succ \boldsymbol{0}
             \end{align}
             The last condition can be relaxed to the following: 
             \begin{align}
              t_{RR,k}'\succeq \epsilon_k \one \label{cont3}
             \end{align}
             where $0<\epsilon_k<1$. Clearly, when we solve problem (\ref{prob})--(\ref{cont3}), this does not mean that the objective function (\ref{prob}) will be zero at this solution. Note further that the optimization problem (\ref{prob})--(\ref{cont3}) is a quadratic convex problem: its objective function is quadratic, and it has a convex equality constraint (\ref{cont1}) and inequality constraints (\ref{cont2}) and (\ref{cont3}). The inequality constraints are element-wise, i.e., $t_{RR,k}'(j)\geq \epsilon_k$ for all $j$, which can be equivalently written as $e_j\tran t_{RR,k}'\geq \epsilon_k$ for all j where $e_j$ is a vector where all its elements are zero expect for the $j^{th}$ element that is one. In this way, the problem becomes a classic constrained convex optimization problem, which can be solved numerically (using for instance interior point methods).

	\section{Simulation Results}
	We illustrate the previous results with the following simulation example. Consider the social network shown in Fig. \ref{network.label.Ex} which consists of $N=23$ agents.
	\begin{figure}[h!]
		\centering
		\includegraphics[scale=0.35]{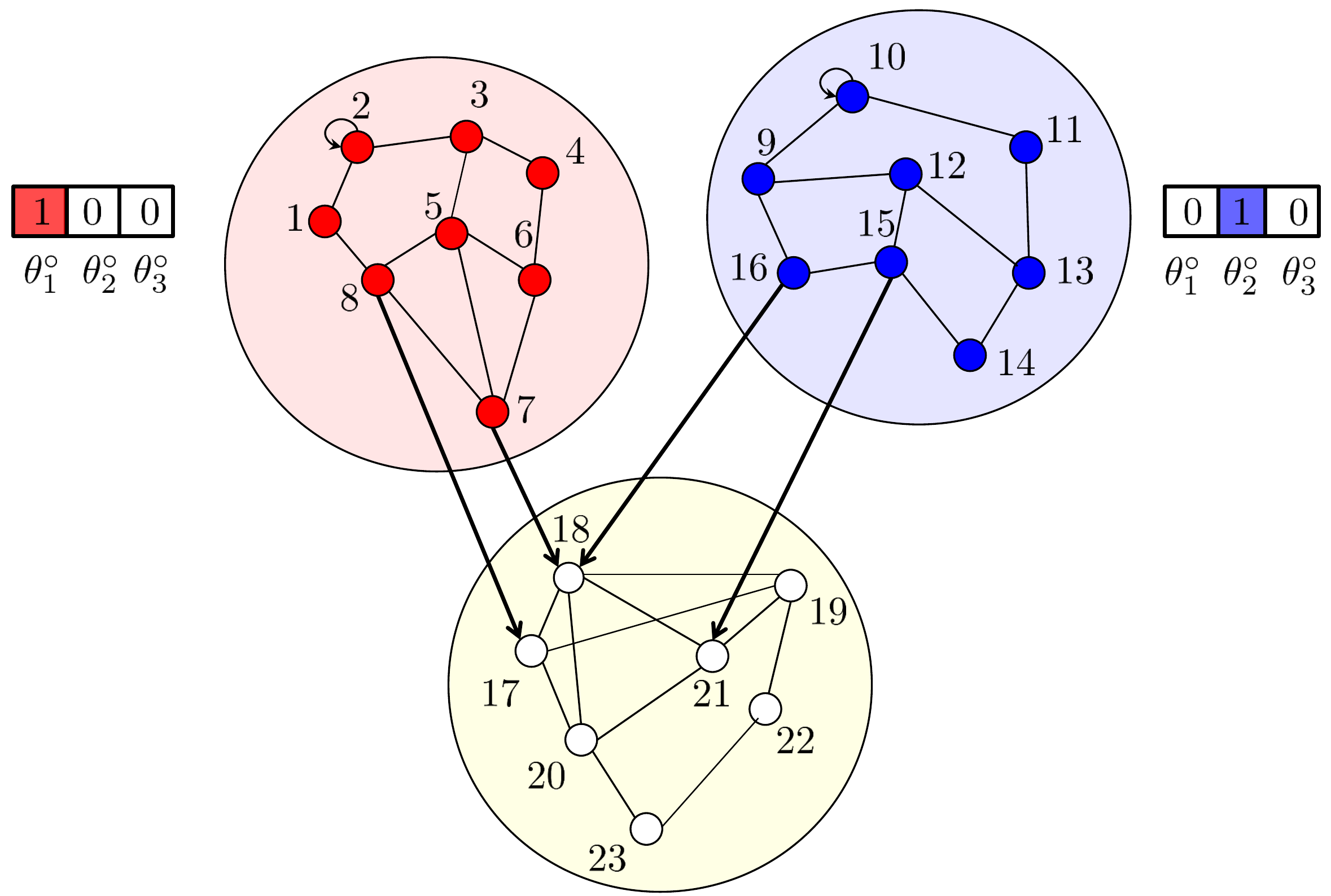} 
		\caption{{\small A weakly-connected network consisting of three sub-networks.}}
		\label{network.label.Ex}
	\end{figure}
	
	\noindent We assume that there are 3 possible events $\Theta=\{\theta_1^\circ,\theta_2^\circ,\theta_3^\circ\}$, where $\theta_1^\circ$ is the true event for the first sending sub-network, $\theta_2^\circ$ is the true event for the second sending sub-network, and $\theta_3^\circ$ is the true event for the receiving sub-network. We further assume that the observational signals of each agent $k$ are binary and belong to $Z_k=\{H,T\}$ where $H$ denotes head and $T$ denotes tail. 
	
		 Agents of the first sending sub-network are connected through the following combination matrix:
		 \begin{align}
		 A_1=\ba{cccccccc}
		 0   &0.3 &0   &0   &0    &0   &0  & 0.3\\
		 0.4 &0  & 0.3 &0   &0    &0   &0  & 0\\
		 0   &0.7& 0   &0.5 &0.25 &0   &0  & 0\\
		 0   &0 &  0.4 &0   &0    &0.3 &0  & 0\\
		 0   &0 &  0.3 &0   &0    &0.1 &0.2& 0.45\\
		 0   &0 &  0   &0.5 &0.25 &0   &0.1& 0\\
		 0   &0 &  0   &0   &0.3  &0.6 &0  & 0.25\\
		 0.6 &0 &  0   &0   &0.2  &0   &0.7& 0
		 \ea
		 \end{align}
		Agents of the second sending sub-network are connected through the following combination matrix:
	 \begin{align}
	 A_2=\ba{cccccccc}
	 0  & 0.35& 0   &0.3& 0   &0&    0&   0.25\\
	 0.1& 0.25& 0.5 &0  & 0  & 0 &   0 &  0\\
	 0  & 0.4 & 0   &0  & 0.8 &0  &  0 &  0\\
	 0.1& 0   & 0   &0  & 0.1 &0   & 0.6& 0\\  
	 0  & 0   & 0.5 &0.3& 0   &0.45 &0  & 0\\
	 0  & 0   & 0   &0  & 0.1 &0    &0.3& 0\\
	 0  & 0   & 0   &0.4& 0   &0.55 &0  & 0.75\\
	 0.8& 0   & 0   &0  & 0   &0    &0.1& 0
	 \ea
	 \end{align}
	The matrices $T_{SR}$ and $T_{RR}$ are going to be designed so that the desired limiting beliefs for receiving agents are as follows:
	\begin{align}
	Q_1=
	\ba{ccccccc}
	0.55&0.5&0.5&0.5&0.45&0.5&0.5\\
	0.45&0.5&0.5&0.5&0.55&0.5&0.5
	\ea \label{Q1}
   \end{align}
   In other words, the weights are going to be designed so that $\theta^\circ_1$ and $\theta^\circ_2$ are almost equally probable for the receiving agents. This illustrates the case when the receiving agents listen to two different perspectives from two media sources that are trustworthy for them, which leaves them undecided regarding which true state to choose. 
   

The likelihood of the head signals for each receiving agent $k$ is selected as the following matrix:
\be
L_{R}(H)
=
\ba{ccccccc}
5/8 & 3/4& 1/6& 7/8& 2/3& 1/3& 1/4 \\
5/8 & 3/4& 1/6& 7/8& 2/3& 1/3& 1/4 \\ 
5/8 & 3/4& 1/6& 7/8& 2/3& 1/3& 1/4 \nn
\ea
\ee 
where each $(j,k)$-th element of this matrix corresponds to $L_k(H/\theta_j)$, i.e., each column corresponds to one agent and each row to one network state. The likelihood of the tail signal is $L(T)=\one_{3\times 7}-L(H)$. The likelihood of the head signals for each sending agent $k$ of the first sending sub-network is selected as the following matrix:
\be
L_{1}(H)
=
\ba{cccccccc}
 5/8 & 3/4 &  1/6 & 1/2 &  1/3  &   1/5 &   4/5 & 1/2 \\  
 5/8 & 3/4 &  1/6 & 2/3 &  1/2  &   1/5 &   2/3 & 1/2 \\  
 1/4 & 3/4 &  1/3 & 1/2 &  1/4  &   1/5 &   4/5 & 1/3
 \ea
\ee 
and the likelihood of the head signals of agents of the second sending sub-network is:
\be
L_{2}(H)
=
\ba{cccccccc}
 7/8       &     5/8  &          1/4   &         1/2     &       1/2      &      1/2 &        6/7   &  1/4\\
 7/8       &     2/3  &          5/8   &         1/3     &       1/2      &      1/2 &         8/9  &   1/4\\
 1/3       &     2/3  &          5/8    &        1/4     &       1/2      &      1/5 &           8/9 &    1/4
\ea
\ee 
\subsection*{Design and Result Simulation}
To achieve $Q_1$, we design $T_{SR}$ and $T_{RR}$ using the results in the previous section. The details of the numerical derivation are omitted for brevity. The non-zero weights in $T_{SR}$ are shown in Fig. \ref{network.label.Ex2}, and $T_{RR}$ is given as follows:
\begin{align}
 T_{RR}
 =\ba{ccccccc}
0    &0.1 & 0.25& 0.25& 0 &   0&    0\\
0.3 &0   & 0.25& 0.25& 0.3 & 0&    0\\
0.3    &0.1 & 0   & 0   & 0.3 & 0.5&  0\\
0.3 &0.1 & 0  &  0   & 0.3&  0  &  0.5\\
0  &0.1 & 0.25& 0.25 &0  &  0  &  0\\
0  &0   & 0.25& 0    &0 &   0  &  0.5\\
0  &0   & 0&    0.25 &0&    0.5 & 0\\
 \ea
 \end{align}
 	\begin{figure}[h!]
 		\centering
 		\includegraphics[scale=0.35]{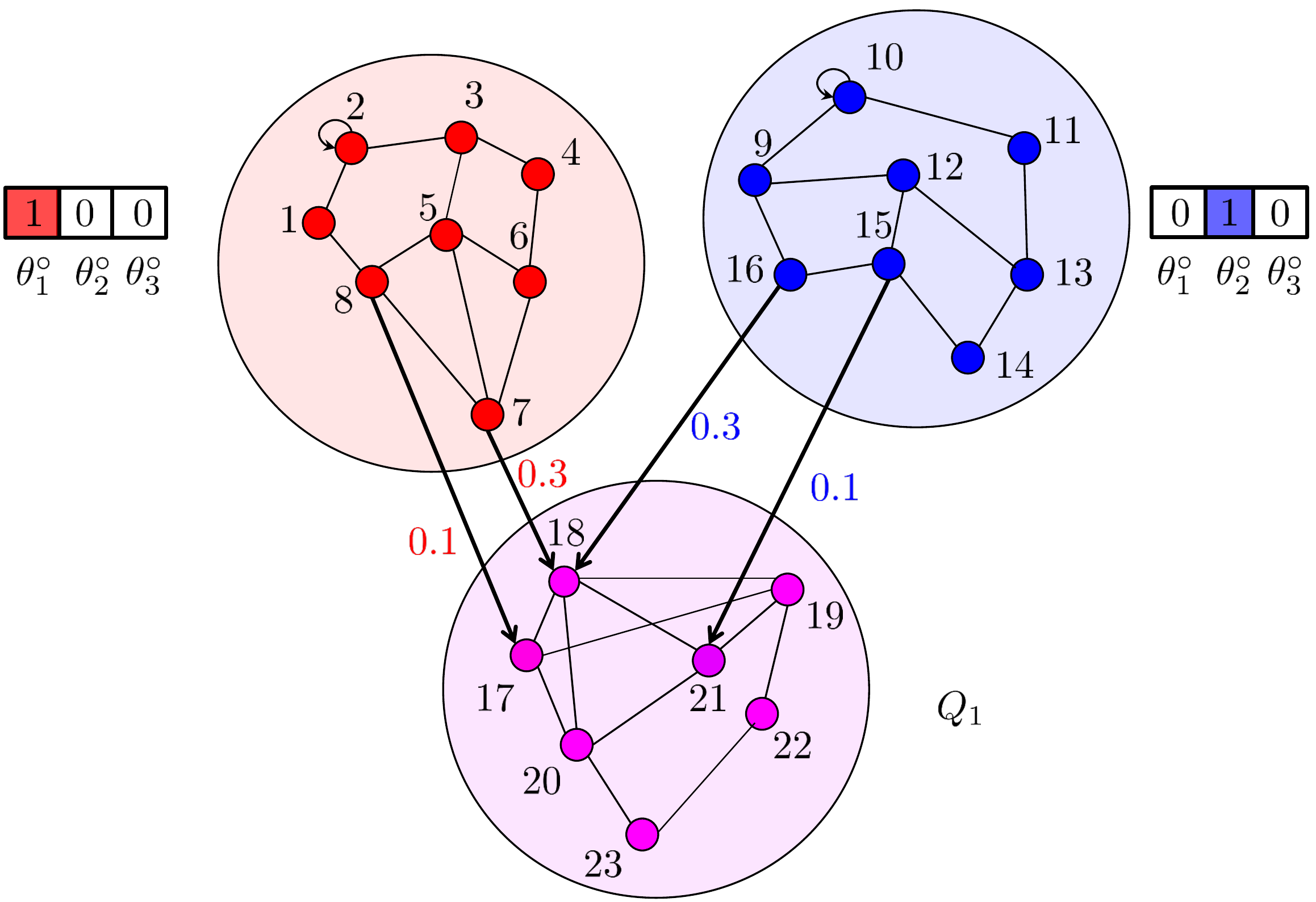} 
 		\caption{{\small Illustration of the limiting beliefs of receiving agents}}
 		\label{network.label.Ex2}
 	\end{figure}
 	
 We run this example for 7000 time iterations. We assigned to each agent an initial belief that is uniform over $\{\theta_1^\circ,\theta_2^\circ,\theta_3^\circ\}$.
 Figures \ref{figTheta1.label} and \ref{figTheta2.label} show the evolution of $\bm\mu_{k,i}(\theta_1^\circ)$ and $\bm\mu_{k,i}(\theta_2^\circ)$ of agents in the receiving sub-network. These figures show the convergence of the beliefs of the agents in the receiving sub-networks to the desired beliefs in $Q_1$.  Figure \ref{network.label.Ex2} illustrates with color the limiting beliefs of receiving agents.
 \begin{figure}[h!]
 	\begin{center}
 		\includegraphics[scale=0.45]{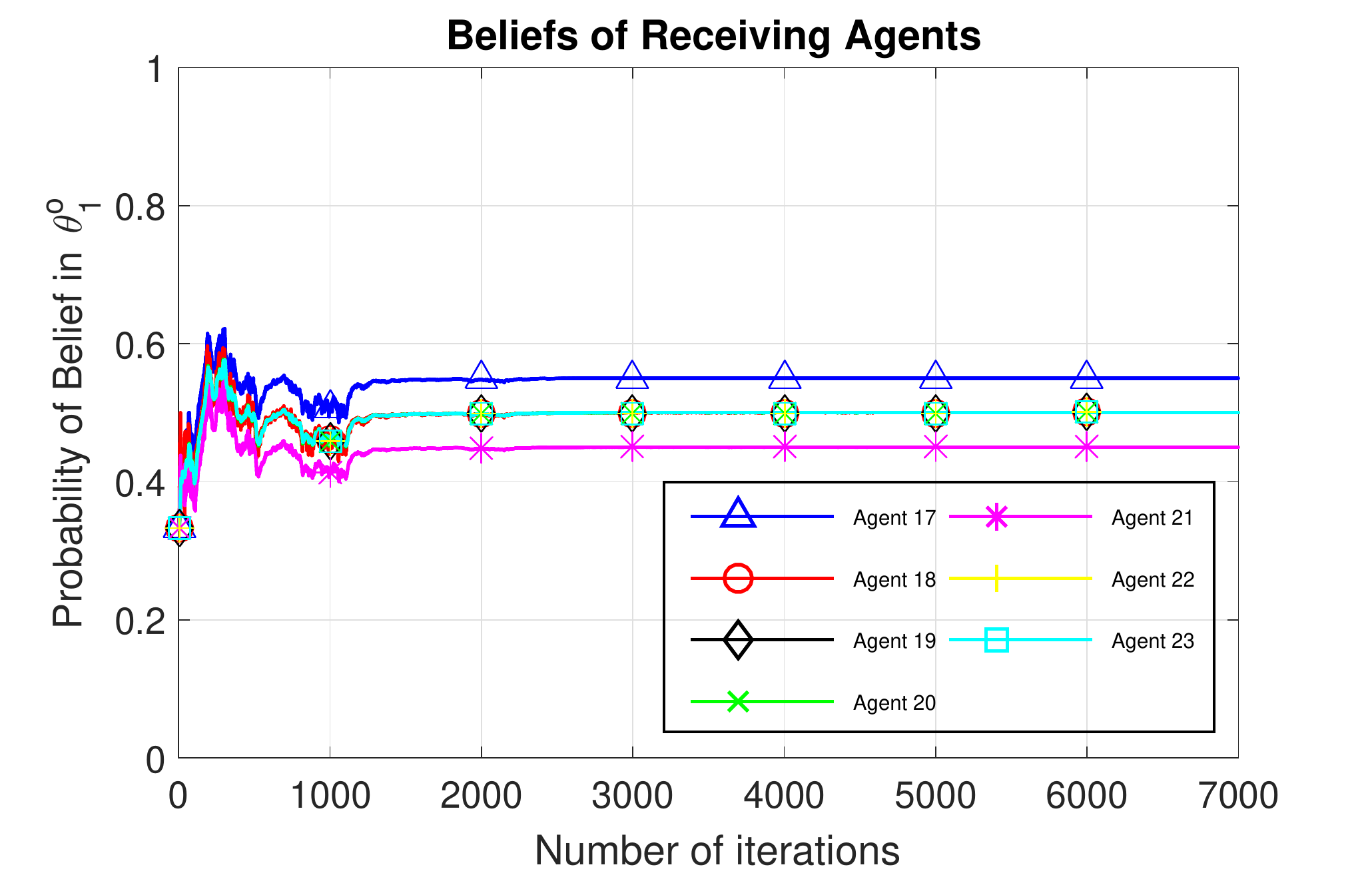} 
 	     \caption{{\small Evolution of the beliefs of the receiving agents at $\theta_1^\circ$ over time}}\label{figTheta1.label}
 	\end{center}
 \end{figure}
 \begin{figure}[h!]

 	\begin{center}
 		\includegraphics[scale=0.45]{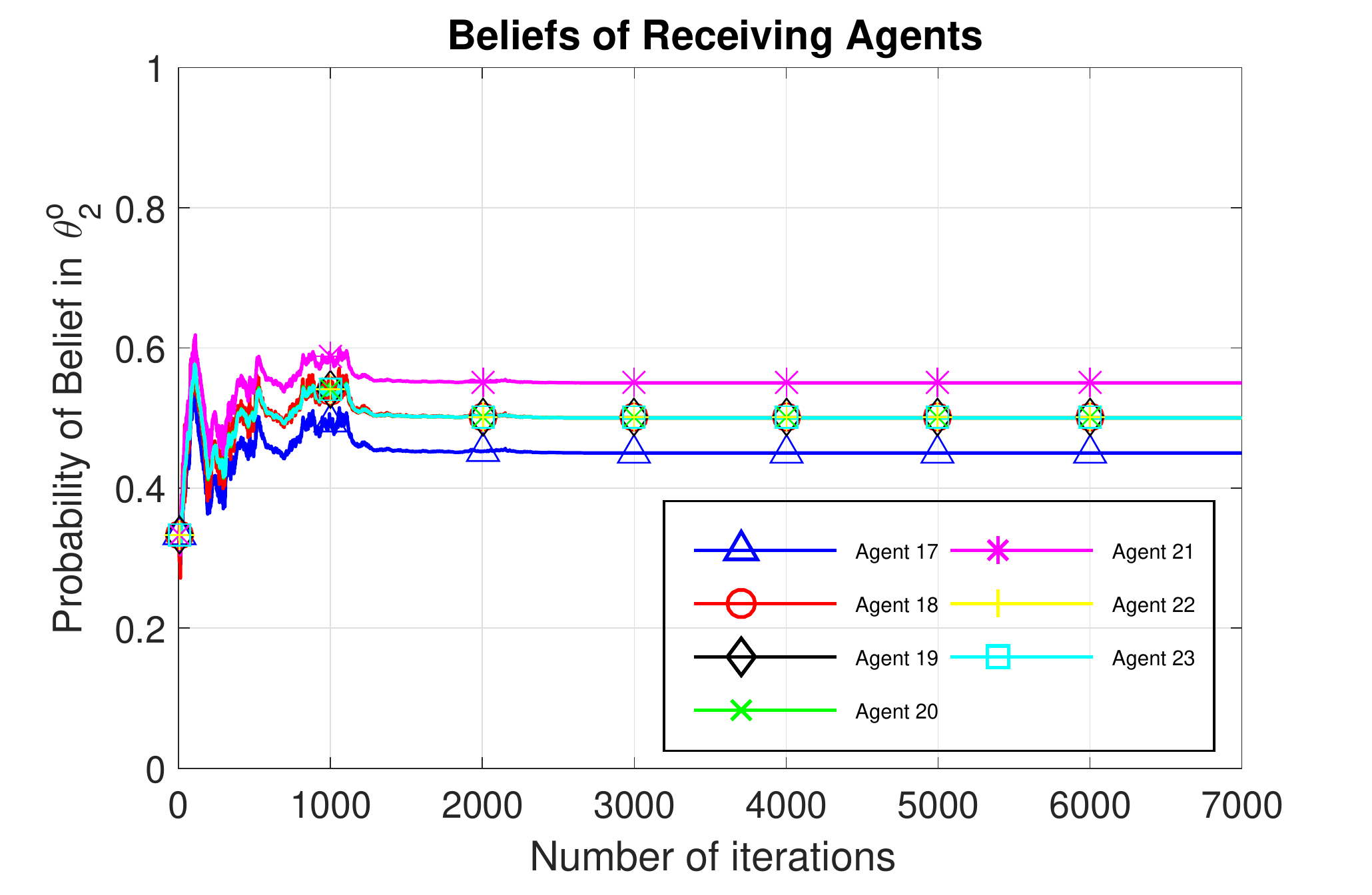}  \caption{{\small Evolution of the beliefs of the receiving agents at $\theta_2^\circ$ over time}}\label{figTheta2.label}
 	\end{center}
 \end{figure}

 \section{Conclusion}
       In this work, we characterized the set of beliefs that can be imposed on non-influential agents and clarified how the graph topology of these latter agents helps resist manipulation but only to a certain degree. We also derived design procedures that allow influential agents to drive the beliefs of non-influential agents to desirable attainable states. 

 \appendices
 \section{Two Revealing Examples for the Design Procedure (\ref{condition2vecR})-(\ref{finalCond3})}\label{App.A}
  \subsection*{Example I: Cases 1 and 2 ($k$ is influenced by sending networks)}
  
  Consider the network shown in Fig. \ref{network.label.whole2}. It consists of $N=8$ agents, two sending sub-networks and one receiving sub-network, with the following combination matrix:
  \begin{small}
  	\begin{equation}
  	A=\left[
  	\begin{array}{ccccc|ccc}
  	0.2		&	0.2 	     	&0.8		& 0 		&	0 		& \times		&0		&0	 \\
  	0 .5   	&      0.4   		&0.1		& 0 		& 	0		&0	&\times		 &0 \\
  	0.3 		& 	0.4		&0.1		&0		& 	0 		&\times		&0	&\times	\\
  	0          	& 	0  		&0		& 0.4 	& 	0.3 		&\times 	&\times		&0\\
  	0          	& 	0  		&0		& 0.6 	& 	0.7 		& 0	 	&0		&0	\\
  	\hline
  	0          	& 	0  		&0		& 0 		& 	0		& 0	 	&\times		&\times	 \\
  	0		&      0		&0		& 0		& 	0		& \times		&0		&\times	 \\
  	0          	& 	0  		&0		& 0 		& 	0		& \times		&\times		&\times	 \\
  	\end{array}
  	\right]
  	\label{labelEx2}\end{equation}
  \end{small}
  \begin{figure}[h!]
  	\centering
  	\includegraphics[scale=0.35]{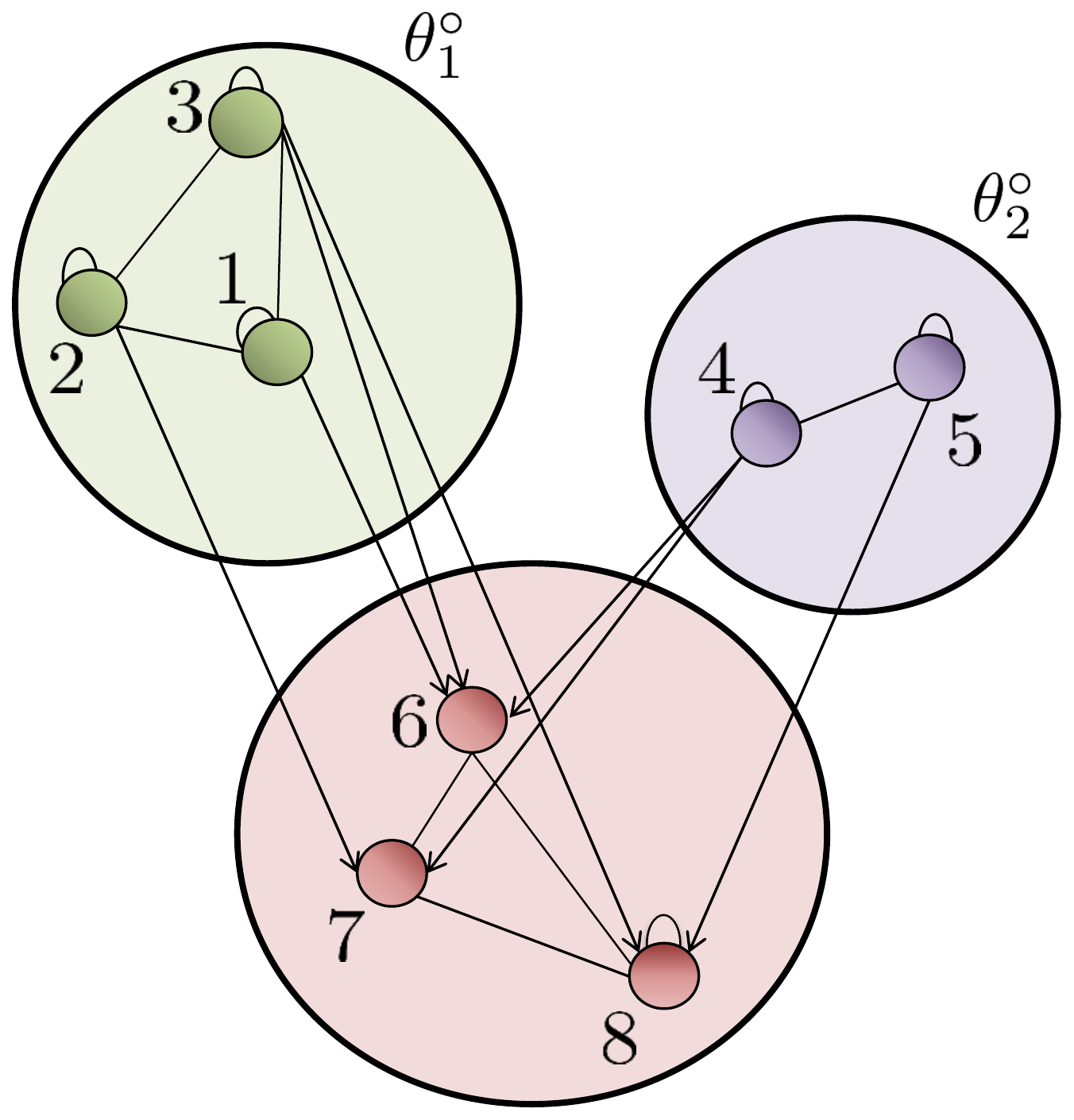} 
  	\caption{{\small A weakly connected network consisting of three sub-networks. In this example, receiving agents $6$ and $7$ are influenced by both sending networks, while agent $8$ is only influenced by the first sending network.}}
  	\label{network.label.whole2}
  \end{figure}
  
  We assume that there are 3 possible states $\Theta=\{\theta_1^\circ,\theta_2^\circ,\theta_3^\circ\}$, where $\theta_1^\circ$ is the true event for the first sending sub-network, $\theta_2^\circ$ is the true event for the second sending sub-network, and $\theta_3^\circ$ is the true event for the receiving sub-network. Let us consider the case where we want to design $T_{SR}$ and $T_{RR}$ to attain the desired limiting beliefs 
  \begin{align}
  Q=\ba{c c c}
  0.2 & 0.3 & 0.5 \\
  0.8 & 0.7 & 0.5
  \ea
  \end{align}
  The matrix $B$ is therefore of the following form:
  \begin{align}
  B=\ba{cccccccc}
  1&1&1&0&0&0.2&0.3&0.5\\
  0&0&0&1&1&0.8&0.7&0.5
  \ea
  \end{align}
  We start with agent $6$. After eliminating entries to satisfy the sparsity in the connections, we are reduced to finding $t_{SR,6}'$ and $t_{RR,6}'$ that satisfy
  \begin{align}
  \underbrace{ \ba{ccccc}
  	1&1&0&0.3&0.5\\
  	0&0&1&0.7&0.5
  	\ea}_{\define B_6}
  \ba{c}
  t_{SR,6}'\\
  t_{RR,6}'
  \ea&=\underbrace{\ba{c} 0.2 \\0.8\ea}_{\define q_6}\label{eq1}
  \end{align}
  Let
  \begin{align}
  \ba{c}
  t_{SR,6}'\\
  t_{RR,6}'
  \ea&\define\ba{ccccc}
  \alpha_1&
  \alpha_2&
  \alpha_3&
  \alpha_4&
  \alpha_5
  \ea\tran\label{form}
  \end{align}
  Agent $6$ is connected to the two sending sub-networks (case 1). Therefore, the problem has a solution, where $t'_{SR,6}$ ($\alpha_1$, $\alpha_2$ and $\alpha_3$) can be expressed in terms of $t'_{RR,6}$ ($\alpha_4$ and $\alpha_5$). More precisely, from (\ref{eq1}) and (\ref{case1}), we have:
  \begin{align}
  \alpha_1+\alpha_2&=0.2-0.3\alpha_4-0.5\alpha_5 \label{cons1}\\
  \alpha_3&=0.8-0.7\alpha_4-0.5\alpha_5 \label{cons2}
  \end{align}
  According to (\ref{case12}), to ensure that $\alpha_1$, $\alpha_2$ and $\alpha_3$ can be chosen as nonnegative numbers, the scalars  $\alpha_4$ and $\alpha_5$ should be chosen to satisfy
  \begin{align}
  0.3\alpha_4+0.5\alpha_5 \leq 0.2 \label{condito1}\\
  0.7\alpha_4+0.5\alpha_5 \leq 0.8 \label{condito2}
  \end{align}
  Note that what matters for scalars $\alpha_1$ and $\alpha_2$ (the weights with which the data received from sending sub-network 1 is scaled) is that their sum should be equal to $0.2-0.3\alpha_4-0.5\alpha_5$ according to (\ref{cons1}). In other words, when a receiving agent is connected to many agents from the same sending sub-network, it does not matter how much weight is given to each of these agents as long as the sum of these weights takes the required value. This is because the beliefs of agents of the same sending sub-networks will converge to the same final distribution. An alternative way to express (\ref{cons1}) is to set $\alpha_1$ and $\alpha_2$ to the following:
  \begin{align}
  \alpha_1&=  \frac{1}{2}\left(0.2-0.3\alpha_4-0.5\alpha_5\right)+\beta \label{k1} \\
  \alpha_2&=  \frac{1}{2}\left(0.2-0.3\alpha_4-0.5\alpha_5\right)-\beta \label{k2}
  \end{align}
  where 
  \begin{align}
  |\beta|\leq \frac{1}{2} (0.2-0.3\alpha_4-0.5\alpha_5)
  \end{align}
  This choice of $\beta$ ensures that $\alpha_1$ and $\alpha_2$ are non-negative and less than $0.2-0.3\alpha_4-0.5\alpha_5$. Moreover, we can check from (\ref{k1}) and (\ref{k2}) that their sum satisfies (\ref{cons1}).
  Therefore, the solution has the following form:
  \begin{align}
  \ba{c}
  t_{SR,6}'\\
  t_{RR,6}'
  \ea&=\ba{c}
  \frac{1}{2}(0.2-0.3\alpha_4-0.5\alpha_5)+\beta\\
  \frac{1}{2}(0.2-0.3\alpha_4-0.5\alpha_5)-\beta\\
  0.8-0.7\alpha_4-0.5\alpha_5\\
  \alpha_4\\
  \alpha_5
  \ea\label{sol67}
  \end{align}
  where
  \begin{align}
  0.3\alpha_4+0.5\alpha_5 \leq 0.2 \label{coni1}\\
  0.7\alpha_4+0.5\alpha_5 \leq 0.8 \label{coni2}\\
  \alpha_4>0,\;\alpha_5 > 0\\
  |\beta|\leq \frac{1}{2}(0.2-0.3\alpha_4-0.5\alpha_5)
  \end{align}
  For example, one solution is to assign the same value $\epsilon_6$ for $\alpha_4$ and $\alpha_5$. Then, from (\ref{coni1}), (\ref{coni2}) and (\ref{epsi1}), we have:
  \begin{align}
  0<\epsilon_6\leq \min \left\{ \frac{0.2}{0.5+0.3}, \frac{0.8}{0.7+0.5}\right\}=0.25 \label{epsi}
  \end{align}
  Let $\epsilon_6=0.1=\alpha_4=\alpha_5$, then
  \begin{align}
  \alpha_1+\alpha_2&=0.2-0.3\alpha_4-0.5\alpha_5 =0.12\\
  \alpha_3&=0.8-0.7\alpha_4-0.5\alpha_5=0.68
  \end{align}
  We can choose $\alpha_1=0.1$ and $\alpha_2=0.02$. Therefore, a possible solution for $t_{SR,6}$ is:
  \begin{align}
  t_{SR,6}=
  \ba{cccccccc}
  0.1&
  0&
  0.02&
  0.68&
  0&
  0&
  0.1&
  0.1
  \ea\tran
  \end{align}
  We follow a similar procedure for agent $7$ and obtain:
  \begin{align}
  \ba{c}
  t_{SR,7}'\\
  t_{RR,7}'
  \ea&\define\ba{c}
  \beta_1\\
  \beta_2\\
  \beta_3\\
  \beta_4
  \ea=\ba{c}
  0.3-0.2\beta_3-0.5\beta_4\\
  0.7-0.8\beta_3-0.5\beta_4\\
  \beta_3\\
  \beta_4
  \ea
  \end{align}
  where 
  \begin{align}
  0.2\beta_3+0.5\beta_4 \leq 0.3 \label{cosi1}\\
  0.8\beta_3+0.5\beta_4 \leq 0.7\label{cosi2}\\
  \beta_3>0,\;\beta_4 >0
  \end{align}
  For this agent, we can choose for instance as a solution $\beta_3=0.2$ and $\beta_4=0.1$ (as they both satisfy (\ref{cosi1}) and (\ref{cosi2})). In this case,
  \begin{align}
  \beta_1&= 0.3-0.2\beta_3-0.5\beta_4=0.21\\
  \beta_2&= 0.7-0.8\beta_3-0.5\beta_4=0.49
  \end{align}
  Therefore, a possible solution for $t_{SR,7}$ is:
  \begin{align}
  t_{SR,7}=
  \ba{cccccccc}
  0&
  0.21&
  0&
  0.49&
  0&
  0.2&
  0&
  0.1
  \ea\tran
  \end{align}
  
  Agent $8$ is connected to the first sending sub-network only (case 2). For this agent, we have:
  \begin{align}
  \underbrace{ \ba{ccccc}
  	1&0.2&0.3&0.5\\
  	0&0.8&0.7&0.5
  	\ea}_{\define B_8}
  \ba{c}
  t_{SR,8}'\\
  t_{RR,8}'
  \ea&=\underbrace{\ba{c} 0.5 \\0.5\ea}_{\define q_8}\label{eq2}
  \end{align}
  Let
  \begin{align}
  \ba{c}
  t_{SR,8}'\\
  t_{RR,8}'
  \ea\define\ba{c}
  \gamma_1\\
  \gamma_2\\
  \gamma_3\\
  \gamma_4
  \ea
  \end{align}
  Therefore, from (\ref{eq2}), (\ref{case2}) and (\ref{case21}), we have:
  \begin{align}
  \gamma_1=  0.5-0.2\gamma_2-0.3\gamma_3-0.5\gamma_4\\
  0.8\gamma_2+0.7\gamma_3+0.5\gamma_4=0.5 \label{cond2Ex}
  \end{align}
  and any vector that satisfies (\ref{eq2}) has the following form:
  \begin{align}
  \ba{c}
  t_{SR,8}'\\
  t_{RR,8}'
  \ea
  &= \ba{c}
  \gamma_1\\
  \gamma_2\\
  \gamma_3\\
  \gamma_4
  \ea
  =\ba{c}
  0.5-0.2\gamma_2-0.3\gamma_3-0.5\gamma_4\\
  \gamma_2\\
  \gamma_3\\
  \gamma_4
  \ea\label{form2}
  \end{align}
  where
  \begin{align}
  0.8\gamma_2+0.7\gamma_3+0.5\gamma_4=0.5
  \end{align}
  Now to ensure that $\gamma_1$ is non-negative, $\gamma_2$, $\gamma_3$ and $\gamma_4$ should be chosen as follows (as in (\ref{case22})):
  \begin{align}
  0.2\gamma_2+0.3\gamma_3+0.5\gamma_4\leq 0.5 \label{cond1Ex}
  \end{align}
  Therefore, a solution in this case should satisfy (\ref{form2}) subject to
  \begin{align}
  0.8\gamma_2+0.7\gamma_3+0.5\gamma_4=0.5\label{c1}\\
  0.2\gamma_2+0.3\gamma_3+0.5\gamma_4\leq 0.5\label{c2}\\
  \gamma_2>0,\; \gamma_3>0,\; \gamma_4> 0\label{c3}
  \end{align}
  For this example, finding $\gamma_2$, $\gamma_3$ and $\gamma_4$ that satisfy 
  (\ref{c1})-(\ref{c2}) is always possible. To see this, for any choice of
  $\gamma_2$, $\gamma_3$ and $\gamma_4$ that satisfy (\ref{c1}), condition (\ref{c2}) is automatically satisfied. Indeed, if (\ref{c1}) is satisfied then
  \begin{align}
  0.5\gamma_2+0.5\gamma_3+0.5\gamma_4&\leq 0.8\gamma_2+0.7\gamma_3+0.5\gamma_4=0.5\\
  \implies 0.5(\gamma_2+\gamma_3+\gamma_4)&\leq 0.5\implies \gamma_2+\gamma_3+\gamma_4\leq 1
  \end{align}
  Therefore,
  \begin{align}
  \gamma_2+\gamma_3+\gamma_4-0.8\gamma_2-0.7\gamma_3-0.5\gamma_4\leq 1- 0.5\\
  \implies 0.2\gamma_2+0.3\gamma_3+0.5\gamma_4\leq  0.5
  \end{align}
  For instance, one possible choice for $\gamma_2$, $\gamma_3$ and $\gamma_4$ that satisfies (\ref{c1}) is 
  \begin{align}
  \gamma_2=\gamma_3=\gamma_4=\frac{0.5}{0.8+0.7+0.5}=0.25
  \end{align}
  Then, 
  \begin{align}
  \gamma_1=0.5-0.2\gamma_2-0.3\gamma_3-0.5\gamma_4=0.25
  \end{align}
  Therefore, a possible solution for $t_{SR,8}$ is:
  \begin{align}
  t_{SR,8}=
  \ba{cccccccc}
  0&
  0&
  0.25&
  0&
  0&
  0.25&
  0.25&
  0.25
  \ea\tran
  \end{align}
  Thus, the overall solution is:
  \begin{align}
  \ba{c}
  T_{SR}\\
  \hline
  T_{RR}
  \ea
  =\ba{ccc}
  0.1 & 0 & 0 \\
  0 & 0.21 & 0 \\
  0.02 & 0 & 0.25 \\
  0.68 & 0.49 & 0 \\
  0    & 0 & 0\\
  \hline
  0&0.2&0.25\\
  0.1&0&0.25\\
  0.1&0.1&0.25
  \ea
  \end{align}
  To verify that the beliefs of the receiving agents converge to the desired beliefs, we compute $W^{\sf T}$ from (\ref{defW1}) and use (\ref{FinalDistribution}) to determine the limiting beliefs at $\theta_1^o$ and $\theta_2^o$ at the receiving agents. This calculation gives
  $$\lim_{i\to\infty}\bm \mu_{k,i}(\theta_1^\circ)=\left\{
  \begin{aligned}
  0.1070+0.0310+0.0620=0.2,\ &\quad k=6\\
  0.0258+0.2247+0.0494=0.3,\ &\quad  k=7\\
  0.0443+0.0852+0.3705=0.5,\ &\quad  k=8
  \end{aligned}\right.$$
  and
  $$\lim_{i\to\infty}\bm \mu_{k,i}(\theta_2^\circ)=
  \left\{
  \begin{aligned}
  0.8,\ &\quad k=6\\
  0.7,\ &\quad  k=7\\
  0.5,\ &\quad  k=8
  \end{aligned}\right.$$
  \subsection*{Example II: Case 3 (agent $k$ not influenced by sending networks)}
  
  Consider the network shown in Fig. \ref{network.label.whole}, with the following combination matrix:
  
  \begin{small}
  	\begin{equation}
  	A=\left[
  	\begin{array}{ccccc|ccc}
  	0.2		&	0.2 	     	&0.8		& 0 		&	0 		& \times		&0		&0	 \\
  	0 .5   	&      0.4   		&0.1		& 0 		& 	0		&0	&\times		 &0 \\
  	0.3 		& 	0.4		&0.1		&0		& 	0 		&\times		&0	&0	\\
  	0          	& 	0  		&0		& 0.4 	& 	0.3 		&\times 	&\times		&0\\
  	0          	& 	0  		&0		& 0.6 	& 	0.7 		& 0	 	&0		&0	\\
  	\hline
  	0          	& 	0  		&0		& 0 		& 	0		& 0	 	&\times		&\times	 \\
  	0		&      0		&0		& 0		& 	0		& \times		&0		&\times	 \\
  	0          	& 	0  		&0		& 0 		& 	0		& \times		&\times		&0	 \\
  	\end{array}
  	\right]
  	\label{labelEx1}\end{equation}
  \end{small}
  \begin{figure}[h!]
  	\centering
  	\includegraphics[scale=0.35]{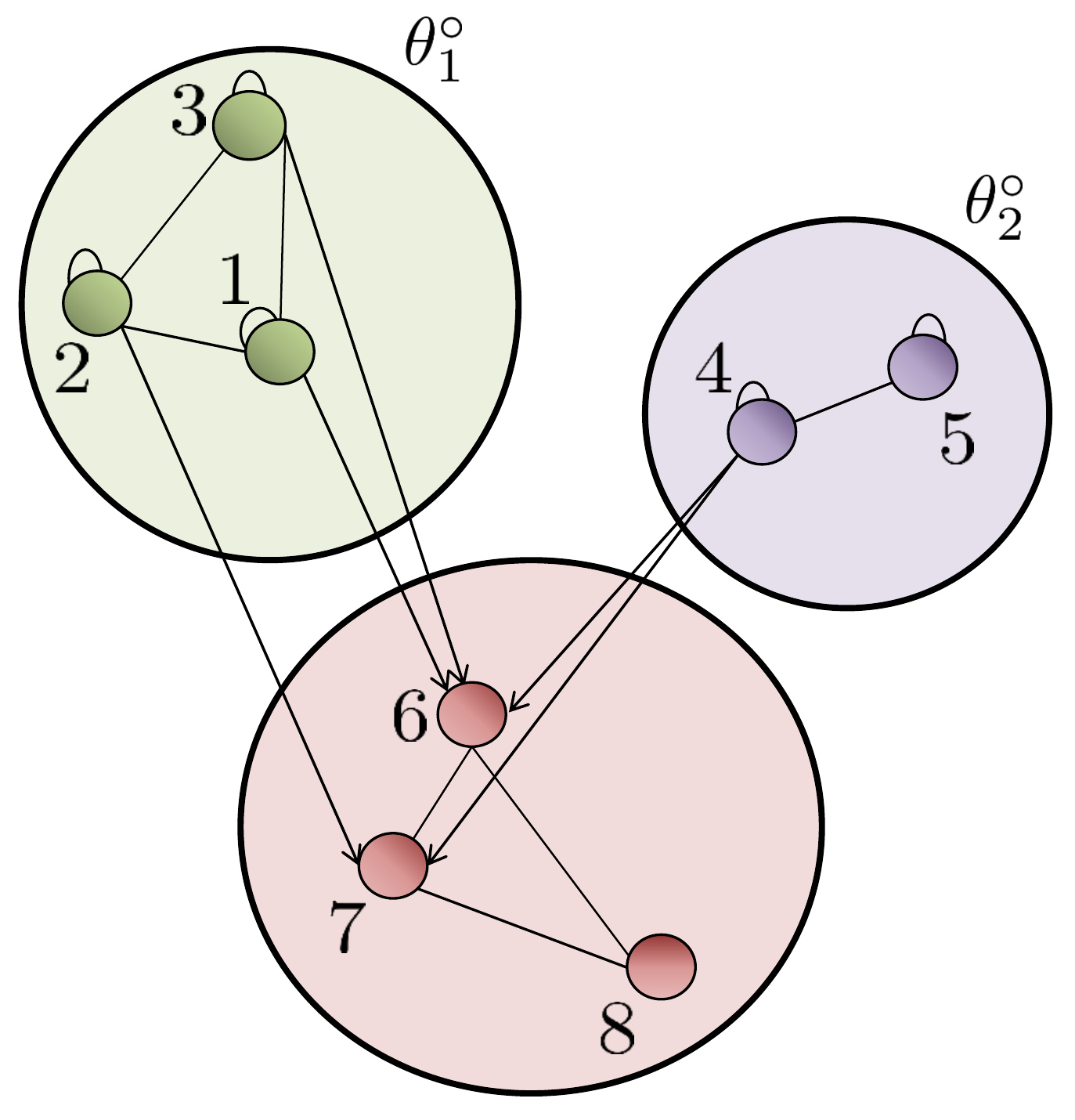} 
  	\caption{{\small A weakly connected network consisting of three sub-networks. In this case, agent $8$ is not influenced by any sending network.}}
  	\label{network.label.whole}
  \end{figure}
  
  \noindent What is different now is that agent $8$ does not have is not connected to agent $3$ (that is, agent $8$ is not connected to any sending network). We are still assuming in this example that we have the same desired limiting beliefs:
  \begin{align}
  Q=\ba{c c c}
  0.2 & 0.3 & 0.5 \\
  0.8 & 0.7 & 0.5
  \ea
  \end{align}
  For agents $6$ and $7$, the solutions found for their corresponding columns are still valid here. However, in this example, $t_{SR,8}$ should have all its elements equal to zero and $t_{RR,8}$ should have its third element equal to zero. Therefore, for agent $8$, the problem reduces to finding $t_{RR,8}'$ that satisfies the following relationship:
  \begin{align}
  \underbrace{ \ba{cc}
  	0.2&0.3\\
  	0.8&0.7
  	\ea}_{\define B_8} t_{RR,8}'&=\underbrace{\ba{c} 0.5 \\0.5\ea}_{\define q_8}
  \end{align}
  where the elements of $t_{RR,8}'$ should be positive and add up to $1$. Any convex combination of $0.2$ and $0.3$ can only produce a number between $0.2$ and $0.3$, but not $0.5$. This is why in this case, the problem does not have a solution. However, we can seek instead a least-squares solution for agent $8$:
  \begin{align}
  \min_{t_{RR,8}'}\left\|B_8 t_{RR,8}'-q_8 \right\|^2 \label{lsq1}
  \end{align}
  subject to
  \begin{align}
  t_{RR,8}'\succeq \epsilon_8 \one\\
  \one\tran t_{RR,8}'=1
  \end{align}
  By choosing $\epsilon_8=0.01$ and solving it numerically, we obtain:
  \begin{align}
  t_{RR,8}'=\ba{c} 0.01\\0.99\ea \label{lsq1sol}
  \end{align}
  This solution can be also deduced directly because $[0.3;0.7]$ is the closer distribution to $[0.5;0.5]$ than any other distribution formed by a convex combination of $[0.2;0.8]$ and $[0.3;0.7]$. Because the entries should be strictly greater than 0, the lowest possible value is given to the first entry of $t'_{RR,8}$. Therefore, with this choice: 
  \begin{align}
  \ba{c}
  T_{SR}\\
  \hline
  T_{RR}
  \ea
  =\ba{ccc}
  0.1 & 0 & 0 \\
  0 & 0.21 & 0 \\
  0.02 & 0 & 0 \\
  0.68 & 0.49 & 0 \\
  0    & 0 & 0\\
  \hline
  0&0.2&0.01\\
  0.1&0&0.99\\
  0.1&0.1&0
  \ea
  \end{align}
  we verify the limiting beliefs of the agents as follows. We compute $W^{\sf T}$ from (\ref{defW1}) and use (\ref{FinalDistribution}) to determine the limiting beliefs at $\theta_1^o$ and $\theta_2^o$ at the receiving agents . This calculation gives
  $$\lim_{i\to\infty}\bm \mu_{k,i}(\theta_1^\circ)=\left\{
  \begin{aligned}
  0.174,\ &\quad k=6\\
  0.272,\ &\quad  k=7\\
  0.271,\ &\quad  k=8
  \end{aligned}\right.$$
  and
  $$\lim_{i\to\infty}\bm \mu_{k,i}(\theta_2^\circ)=
  \left\{
  \begin{aligned}
  0.826,\ &\quad k=6\\
  0.728,\ &\quad  k=7\\
  0.729,\ &\quad  k=8
  \end{aligned}\right.$$
  
  It is expected that the beliefs of agents $6$ and $7$ would not converge to the desired beliefs, because the belief of agent $8$ cannot converge to its desired belief, which will definitely affect the limiting beliefs of agents $6$ and $7$. We know that agent $8$ will not converge to its desired limiting belief because [0.5;0.5] cannot be obtained by any convex combination of [0.2;0.8] and [0.3;0.7] (its neighbors' limiting beliefs, (\ref{roula})). 

  	  \bibliographystyle{IEEEbib}

\end{document}